%% file: main.tex
\documentclass[sigconf]{acmart}

\AtBeginDocument{%
  \providecommand\BibTeX{{%
    \normalfont B\kern-0.5em{\scshape i\kern-0.25em b}\kern-0.8em\TeX}}}

\setcopyright{acmlicensed}
\copyrightyear{2025}
\acmYear{2025}
\acmDOI{XXXXXXX.XXXXXXX}

\acmConference[CHI '25]{Proceedings of the 2025 CHI Conference on Human Factors in Computing Systems}{April 26-- May 1, 2025}{Yokohama, Japan}
\acmISBN{978-1-4503-XXXX-X/18/06}

\usepackage{enumitem}
\usepackage{natbib}
\usepackage{tabularx}
\usepackage{subcaption} 

\usepackage{framed}
\copyrightyear{2026}
\acmYear{2026}
\setcopyright{cc}
\setcctype{by}
\acmConference[CHI '26]{Proceedings of the 2026 CHI Conference on Human Factors in Computing Systems}{April 13--17, 2026}{Barcelona, Spain}
\acmBooktitle{Proceedings of the 2026 CHI Conference on Human Factors in Computing Systems (CHI '26), April 13--17, 2026, Barcelona, Spain}
\acmPrice{}
\acmDOI{10.1145/3772318.3790775}
\acmISBN{979-8-4007-2278-3/2026/04}

\begin{document}

\title[Belief Updating and Delegation in Multi-Task Human–AI Interaction: Evidence from Controlled Simulations]{Belief Updating and Delegation in Multi-Task Human–AI Interaction: Evidence from Controlled Simulations}





\author{Shreyan Biswas}
\affiliation{%
  \institution{Delft University of Technology}
  \city{Delft}
  \country{The Netherlands}}
\email{s.biswas@tudelft.org}

\author{Alexander Erlei}
\affiliation{%
  \institution{University of Goettingen}
  \city{Goettingen}
  \country{Germany}}
\email{alexander.erlei@wiwi.uni-goettingen.de}

\author{Ujwal Gadiraju}
\affiliation{%
  \institution{Delft University of Technology}
  \city{Delft}
  \country{The Netherlands}}
\email{u.k.gadiraju@tudelft.nl}

\begin{abstract}

Large language models (LLMs) increasingly support heterogeneous tasks within a single interface, requiring users to form, update, and act upon beliefs about one system across domains with different reliability profiles. Understanding how such beliefs transfer across tasks and shape delegation is therefore critical for the design of multipurpose AI systems. We report a preregistered experiment ($N=240$, 7,200 trials) in which participants interacted with a controlled AI simulation across grammar checking, travel planning, and visual question answering, each with fixed, domain-typical accuracy levels. Delegation was operationalized as a binary reliance decision—accepting the AI’s output versus acting independently and belief dynamics were evaluated against Bayesian benchmarks. We find three main results. First, participants do not reset beliefs between tasks: priors in a new task depend on posteriors from the previous task, with a 10-point increase predicting a 3–4 point higher subsequent prior. Second, within tasks, belief updating follows the Bayesian direction but is substantially conservative, proceeding at roughly half the normative Bayesian rate. Third, delegation is driven primarily by subjective beliefs about AI accuracy rather than self-confidence, though confidence independently reduces reliance when beliefs are held constant. Together, these findings show that users form global, path-dependent expectations about multipurpose AI systems, update them conservatively, and rely on AI primarily based on subjective beliefs rather than objective performance. We discuss implications for expectation calibration, reliance design, and the risks of belief spillovers in deployed LLM-based interfaces.

\end{abstract}


\maketitle
\input{tex/1_Introduction}

\input{tex/2_Background}
\input{tex/3_Experiment_Design}
\input{tex/4_Analysis}

\input{tex/5_Results}
\input{tex/6_Discussiom}

\input{tex/7_Conclusion}

\begin{acks}
We thank all the anonymous participants in our study. This work was partially supported by the TU Delft AI Initiative, the Model Driven Decisions Lab (\textit{MoDDL}), the Robust LTP GENIUS Lab, and the \textit{ProtectMe} Convergence Flagship. Some sentences were refined for clarity using ChatGPT. LLMs also assisted in generating utility code (e.g., web app scaffolding, analysis scripts, plotting), which were reviewed and validated by the authors. 
\end{acks}

\bibliographystyle{ACM-Reference-Format}
\bibliography{main}
\input{tex/8_Appendix}

\end{document}

%% file: tex/1_Introduction.tex
\section{Introduction}


It is not hard to imagine a student who uses AI to fix grammar in an essay, then without hesitation trusts the same system to recommend a travel itinerary for an unfamiliar city. Today’s multipurpose AI systems invite exactly this kind of cross-domain reliance—the same AI can edit text, plan a trip, and answer knowledge-intensive questions within a single conversation \cite{handa2025economictasksperformedai}. Yet how people update their beliefs about AI accuracy as they move between such disparate tasks remains relatively underexplored.

Research in the HCI community has long studied how people form and calibrate trust in automation \cite{parasuraman2000model}. Early work in aviation \cite{lee2004trust}, driving \cite{wintersberger2016trust}, and clinical decision support \cite{bussone2015role} demonstrated that both over-trust and under-trust can produce severe risks \cite{dzindolet2003role}. More recent HCI research has extended these insights to AI systems, exploring various models of trust~\cite{mehrotra2024systematic}, testing interventions such as explanations \cite{kahr2024understanding,he2025conversational,robbemond2022understanding}, uncertainty cues \cite{lee2025towards}, and transparency mechanisms \cite{wang2024task}, as well as human factors like mental models~\cite{bansal2019beyond}, cognitive biases~\cite{he2023knowing}, domain experience and expertise~\cite{zhang2022you,nourani2020role}, expectations~\cite{khadpe2020conceptual}, gaze~\cite{cao2022understanding, wu2025releyeance}, and intuition~\cite{chen2023understanding}. The potential of AI systems to support, complement, and augment human capabilities across different tasks has led to widespread applications in human–AI collaborative configurations with varying effectiveness~\cite{wang2020human,fan2022human,vodrahalli2022uncalibrated,reverberi2022experimental,hemmer2023human,ashktorab2020human}. Together, this body of work shows that trust, reliance, and performance in human–AI interaction are shaped by a complex interplay of task properties, system cues, and human cognition—yet typically studied within single-task or single-domain settings, leaving open how users update beliefs and calibrate reliance when the same AI system is used across heterogeneous tasks.

Although prior work has explored various task factors, human factors, and system factors that influence and shape experiential and performance-related outcomes in human-AI collaboration, much of the existing literature has considered the context of user interactions with a single AI system to accomplish a given task.
Such work has contributed to enriching our understanding of these factors, but does not sufficiently reflect the multi-task and multipurpose interactions with AI systems prevalent in today's world. While in the former case, users can incrementally learn about the reliability of a given AI system and adapt their behavior, in the latter case of interactions with a multipurpose AI system where accuracy varies across tasks, this assumption breaks down. 

We aim to address this crucial gap in existing research and in the current understanding of how users update their beliefs about a multipurpose AI system, as they interact with the system to accomplish different tasks.

We can expect that rational users would reset their beliefs about multipurpose AI systems when accomplishing a new task. This is particularly due to potential variability in the performance of such systems across different tasks. For instance, in today's world, the same large language models (LLMs) are used across domains to accomplish different tasks---from writing to planning, information search, decision-making, and analysis~\cite{brachman2024knowledge, lee2025impact, handa2025economictasksperformedai}. 
It is important to understand how people carry over their expectations across tasks and whether they reset their beliefs for each new task that they accomplish with an  AI system.

We aim to address the following research questions:
\begin{framed}
    \begin{itemize}[leftmargin=*]
        \item \textbf{RQ1:} How do users form, update, and transfer beliefs about an AI system’s accuracy when interacting with the same system across multiple tasks with different accuracies?
        \item \textbf{RQ2:} How do users’ beliefs about AI accuracy and their self-confidence jointly influence delegation decisions in human–AI interaction?
        \item \textbf{RQ3:} How do dispositional trust and related individual differences shape users’ initial beliefs about AI accuracy before observing performance?
        
    \end{itemize}
\end{framed}

To this end, we conducted a sequential, multi-task experiment in which participants were asked to complete three distinct real-world tasks with different accuracies in the presence of AI-assistance: 
grammar correction, visual question answering, and travel planning.
We employed pre-scripted AI outputs to simulate controlled accuracy levels, enabling precise tests of belief updating and delegation. Guided by prior literature---elucidated in the next section---on trust calibration, belief updating, and reliance in human–automation interaction, we preregistered the following hypotheses:

\begin{itemize}[leftmargin=*]
    \item \textbf{H1 (Cross-task priors):} Priors about AI accuracy in a new task will be anchored to posterior beliefs from the preceding task rather than fully resetting (RQ1).  
    \item \textbf{H2 (Bayesian vs.\ bounded updating):} Within tasks, participants will update their beliefs in the Bayesian-predicted direction but less strongly, consistent with conservatism bias (RQ1).  
    \item \textbf{H3 (Beliefs $\to$ Delegation):} Higher lagged beliefs (i.e., beliefs from previous task) about AI accuracy will increase the likelihood of delegating to the AI (RQ2).  
    \item \textbf{H4 (Confidence $\to$ Delegation):} Higher self-confidence will decrease delegation, holding beliefs constant (RQ2).  
    \item \textbf{H5 (Dispositional trust $\to$ Priors):} Individuals with higher propensity to trust automation will report higher initial priors about AI accuracy, with AI literacy and need for cognition as potential moderators (RQ3).  
\end{itemize}

We found that priors did not reset across tasks but instead carried over, demonstrating systematic belief inertia (H1). Within tasks, participants updated beliefs in the right direction but only about half as much as Bayesian rationality prescribes, showing a conservatism bias (H2). Delegation was strongly predicted by users’ beliefs about AI accuracy but not by their self-confidence once beliefs were controlled (H3–H4). Finally, dispositional trust reliably predicted higher initial priors, with AI literacy providing an independent boost (H5).

This paper contributes both empirical evidence and theoretical insight. To the best of our knowledge, we provide the first systematic evidence of belief updating and delegation in a multi-task, multi-accuracy human–AI setting. We show that rather than resetting as rational agents would, users carry forward beliefs across tasks, linking experience in one domain to expectations in another through boundedly rational updating. These findings carry direct implications for the design of multipurpose AI systems, where cross-task reliance is the norm rather than the exception.

%% file: tex/2_Background.tex
\section{Background and Related Literature}

\subsection{Theoretical Foundations: Bayesian Updating vs. Bounded Rationality}

\paragraph{Bayesian Updating as a Normative Benchmark.}  
In Bayesian epistemology, an ideal rational agent updates beliefs by conditionalization—adjusting prior probabilities in proportion to the likelihood of new evidence. Formally, given a prior belief $P_{\text{old}}(H)$ about a hypothesis $H$ and receiving new evidence $E$, Bayes’ rule prescribes:
$$
P(H) = \frac{P_{\text{old}}(H)\,P(E \mid H)}{P(E)}
$$
This provides a normative benchmark: a rational Bayesian multiplies prior odds by the likelihood ratio of $E$, ensuring path independence. If multiple independent pieces of evidence $E_1, E_2, \dots, E_n$ arrive, it does not matter whether an agent updates sequentially or all at once—the posterior should be identical \cite{benjaminErrorsProbabilisticReasoning2019a}. Our study builds directly on this tension: while Bayesian updating provides the benchmark (\textbf{H2}), we test whether human belief paths in multi-task AI use adhere to this standard or exhibit bounded-rational deviations. In repeated Bernoulli tasks such as ours, Bayesian updating is often formalized using the Beta–Binomial model, where prior beliefs are represented as pseudocounts of imagined past successes and failures \cite{murphy2012machine}. The total defines a ‘stickiness’ parameter controlling conservatism: large values resist change, small values amplify responsiveness.

\paragraph{Systematic Deviations and Biases.}  
Decades of psychological and behavioral economics research show that human updating often departs from this benchmark. Conservatism, or under-reaction, describes the tendency to adjust beliefs less than Bayes’ rule predicts \cite{Edwards1968,kahneman1972subjective,Jakobsen2020CoarseBU}. Over-reaction, by contrast, occurs when people overweight salient or recent information, a pattern sometimes described as “seeing patterns in noise.” Other well-documented biases—such as anchoring \cite{Tversky1974}, belief inertia, and confirmation bias \cite{rabin1999first,charness2005optimal}—further highlight the bounded rationality of human updating. These findings suggest that while Bayesian updating provides a gold standard, real-world human cognition often follows heuristics or simplified mental models. Understanding how humans update their beliefs in a multipurpose AI system with varying accuracy across tasks is both a timely and important question to explore.

While the distinctions above clarify normative versus descriptive accounts, several mechanism-level models from behavioral economics and cognitive psychology further specify how bounded rationality arises. Noisy-sampling models propose that people approximate likelihoods with limited internal samples, producing attenuated updates that resemble conservatism \cite{benjaminErrorsProbabilisticReasoning2019a}. Inattention models \cite{gabaix2014sparsity} assume that individuals do not process all available evidence, effectively discounting incoming signals in proportion to cognitive costs. Sequential belief–averaging models such as the Hogarth–Einhorn adjustment rule \cite{hogarth1992order} explain primacy or recency through differential weighting of early versus late information. Confirmation-bias models \cite{rabin1999first} predict asymmetric integration, where evidence consistent with a prior is incorporated more strongly than disconfirming information. Incorporating these models allows us to interpret deviations from Bayesian predictions in mechanistic terms—e.g., whether under-reaction in our study reflects cognitive load, asymmetric weighting, or simple linear updating heuristics.

\subsection{Belief Updating in Sequential Contexts}

Order effects illustrate how human updating departs from Bayesian path independence. Classic work on belief adjustment by \citet{hogarth1992order} shows that the sequence of evidence can yield primacy (overweighting early evidence) or recency (overweighting later evidence) effects. More recent studies confirm that anchors and temporal biases distort sequential reasoning in both human judgment and human–AI interaction \cite{schreiner2024time,echterhoff2022ai}. For instance, \citet{schreiner2024time} found that when people evaluate sequential evidence, they often adopt a "weight-last-stronger strategy - placing disproportionate weight on the most recent outcome. Such recency effects imply that sequential interaction with AI may elicit boundedly rational belief dynamics rather than normative Bayesian updating. In our experiment, we leverage sequential feedback in a human-ai collaboration setup to test whether belief changes trace a normative (Beta–Binomial) step or show systematic under-reaction or over-reaction.

\subsection{Trust in Automation: Constructs, Calibration, and Delegation}

\paragraph{Layers of trust.}  
Trust in automation—a user’s confidence that a system will perform as intended—has been organized into dispositional, situational, and learned components \cite{hoff2015trust}. Meta-analyses and reviews consistently identify perceived reliability/accuracy as a primary antecedent of trust and reliance \cite{schaefer2016meta,glikson2020human}. In short, when users judge an AI to be more accurate, it is rational (and empirically common) to trust and use it more.

\paragraph{Belief calibration about AI performance.}  
Beliefs about an AI’s accuracy are updated through experience: observed successes raise perceived reliability; observed failures reduce it. Yet calibration is imperfect and often lags true performance \cite{dzindolet2003role,hutchinson2022human,stricklandHowHumansLearn2024,he2023stated}. Trial-by-trial studies find recency-weighted belief paths and incomplete convergence within typical experimental windows, motivating designs that make performance signals clearer and more diagnostic. Our study extends this line of work by embedding calibration into three distinct tasks with fixed accuracy levels, enabling direct tests of Bayesian vs. bounded-rational updating (\textbf{H2}).

\paragraph{From belief to delegation.}  
Trust matters because it shapes delegation: whether users accept or reject AI recommendations. Higher perceived reliability is associated with greater advice-taking and reliance \cite{madhavan2007effects,market4lemons2026}. A classic HCI view models delegation as a comparison between trust in the automation and confidence in oneself; when trust exceeds self-confidence, reliance is more likely \cite{lee2004trust,mehrotra2024systematic}. At the same time, “algorithm aversion” shows that a small number of visible errors can sharply depress trust—beyond what Bayesian updating would suggest—reducing reliance even on superior algorithms \cite{dietvorst2015algorithm}. These findings motivate our tests of how lagged beliefs and self-confidence jointly predict delegation across trials and tasks (\textbf{H3, H4}).

It is important to clarify that the form of delegation studied here corresponds to a direct reliance decision commonly considered in contexts of automated decision support systems~\cite{erlei2022s,erlei2020impact, erlei2024understanding}, rather than the multi-turn, mixed-initiative collaboration that characterizes many modern LLM applications. In HCI research, rich interactions such as co-writing, iterative prompting, and stepwise refinement are typically framed as collaborative or co-creative workflows, where users retain continuous oversight and engage in verification and editing behaviors \cite{lee2022coauthor,reza2024abscribe}. Such workflows involve a sequence of micro-decisions rather than a single transfer of control. By contrast, our study isolates the core mechanism of direct delegation: whether a user elects to rely on the AI’s output or act independently, aligning with established paradigms in advice-taking, trust calibration, and one-shot automation reliance \cite{he2023knowing,bansal2021most}. This operationalization allows us to cleanly model the relationship between beliefs, confidence, and reliance without the confounds introduced by iterative editing, exploration strategies, or mixed-initiative negotiation. Such a clean operationalization is necessary for establishing a mechanism-level account of how users integrate evidence into beliefs and reliance decisions. Foundational work on belief adjustment and human–automation reliance has consistently relied on controlled, one-shot paradigms to identify cognitive mechanisms before extending them to richer interactive settings \cite{hogarth1992order,lee2004trust}. Establishing this mechanism-level baseline is therefore a prerequisite for studying more complex, mixed-initiative forms of delegation. Binary delegation is also implicitly embedded in many real-world AI-assisted applications. Code editors such as Visual Studio Code with GitHub Copilot present inline suggestions that users can accept, ignore, or replace—an interaction that directly operationalizes reliance as an accept–reject decision. Similarly, writing assistants (e.g., Google Docs’ Smart Compose, Microsoft Editor, Grammarly) surface sentence- or paragraph-level suggestions that users adopt or dismiss with a single action. Even when these systems support multi-step refinement, the fundamental unit of interaction is an accept–reject decision on specific AI contributions. These patterns demonstrate that binary delegation is not only a controlled experimental abstraction but a common and ecologically valid feature of contemporary human–AI workflows.

We also clarify the distinction between delegation and consultation, which is central to accurately situating our operationalization within HCI literature. In HCI, consultation refers to seeking information, suggestions, or inspiration from an AI system while retaining full agency over the final decision—for example, asking an AI for itinerary ideas, code snippets, or text rewrites and then editing or discarding them. Delegation, by contrast, involves transferring decision authority to the system: adopting an AI-proposed answer, accepting an inline code suggestion, or letting the AI's output stand without revision. Our study focuses specifically on this latter form of direct reliance. This distinction aligns our operationalization with established work in advice-taking, trust calibration, and automation reliance, and clarifies that the behaviors we measure correspond to a well-defined and widely used construct rather than general consultation or co-creative interaction.

\paragraph{Dispositional trust and initial priors.} 
Baseline expectations are often shaped by dispositional traits rather than observed evidence. Previous research has shown that a higher propensity to trust automation predicts higher initial trust and earlier reliance—even before performance evidence is available \cite{hoff2015trust, jessup2019measurement}. In our design, this provides the basis for H5: participants with higher trust propensity (TiA) are expected to report more optimistic priors about AI accuracy at the start of each task. Related traits such as AI literacy (MAILS) and need for cognition (NCS-6) were also measured to test whether knowledge or cognitive style similarly color these baseline expectations.

\subsection{Spillovers and Multi-Task Settings}

Beyond single-task paradigms---some of which have considered interactions over time~\cite{tolmeijer2021second,kahr2024understanding,kahr2025good}, recent work examines how beliefs and attitudes about AI spill over across models or domains. For example, \citet{de2023retrospective} show that when individuals integrate multiple model outputs in financial judgment tasks, recency and confirmation biases distort aggregation, leading to systematic deviations from rational integration. \citet{richardson2024one} extend this to algorithmic trust, showing a “one bad apple” effect: after interacting with multiple algorithms across prediction tasks (e.g., forecasts and categorizations), poor performance by one system reduces willingness to use other, unrelated algorithms. They identify human–AI collective efficacy as a mediator of this spillover. Similarly, \citet{Manoli2024} demonstrate a “double standard” in moral evaluation: after observing harmful behavior by one AI in simulated multi-agent tasks, participants generalized distrust and moral concern to all AI systems more strongly than they did for human agents. Together, these studies highlight that spillovers are a robust phenomenon: people often fail to compartmentalize judgments across different systems.

Our work differs in two key respects. First, rather than studying attitudes (trust, aversion, or moral evaluation) that spill over, we focus on beliefs about accuracy—a quantitative construct that can be directly benchmarked against Bayesian rationality. Second, prior spillover studies typically involve distinct systems or hypothetical scenarios (e.g., different algorithms, multiple AI agents). By contrast, our design systematically manipulates ground-truth accuracy across three controlled tasks with the same AI partner. This allows us to isolate whether and how participants carry over accuracy priors across tasks (\textbf{H1}), even when the true reliability changes. In doing so, we bridge the existing literature on spillover effects with formal models of belief updating, testing not only whether spillovers occur in the context of user interactions with multipurpose AI systems but also whether they conform to or deviate from Bayesian predictions.

A consistent theme in prior literature as recently synthesized by \citet{gadiraju2025enterprising} is that users form global expectations or mental models of an AI system that shape how new information is interpreted. \citet{bansal2019beyond} found that human mental models of AI capabilities and how they base their mental models to accept or override an AI system’s recommendation can affect human-AI team
performance. Work by \citet{kocielnik2019will} shows that expectations about imperfect AI strongly influence satisfaction, perceived usefulness, and eventual acceptance. 
Prior work by \citet{khadpe2020conceptual} found that conceptual metaphors used to describe AI systems can shape user expectations and behavior in human-AI collaboration. In conversational crowdsourcing, \citet{jung2022great} found that  different human and non-human
metaphors can affect user engagement, their perceived cognitive load in various tasks, their intrinsic motivation, and their trust in the agents.
In human–automation research, similar dynamics appear as the reliance gap, where users overweight visible failures relative to successes \cite{dzindolet2003role}, and as misuse or disuse patterns when mental models do not match actual system capabilities \cite{lee2004trust}. \citet{nourani2022importance} identified the role of experiential biases in shaping user mental models and outcomes in their interactions with AI systems. First impressions of AI systems have been shown to influence user expectations and reliance behavior on AI systems~\cite{nourani2020investigating}, although the evolution of user perceptions beyond first impressions have also been identified in different contexts~\cite{tolmeijer2021second}.

Our belief-updating framework complements this prior work by providing a mechanistic account of how such mental models evolve over time. Rather than measuring expectations only as static ratings, we track the dynamics of belief change—revealing how anchoring \cite{tversky1974judgment,nourani2021anchoring}, under-reaction \cite{edwards1968conservatism, benjaminErrorsProbabilisticReasoning2019a}, and cross-task spillovers \cite{biswas2025mind, richardson2024one, de2023retrospective} influence users’ perceptions of a multipurpose AI. This positions belief updating as a process-level construct that links these cognitive biases with well-established HCI phenomena such as trust calibration \cite{hoff2015trust, glikson2020human}, users’ error-management and verification strategies \cite{smith2020no,bansal2019beyond}, and broader patterns of miscalibration, over-reliance, and disuse documented in human–automation research \cite{lee2004trust,dzindolet2003role}.

%% file: tex/3_Experiment_Design.tex
\section{Study Design}

Our study comprises a single experiment that received approval from our institutional ethics review committee. Participants interacted with a simulated multipurpose AI assistant across three different tasks:  

\begin{itemize}[leftmargin=*]
    \item \textbf{Grammar Error Detection:} Sentences were sampled from the CoNLL-2014 shared task on grammatical error correction \cite{ng-etal-2014-conll}. Participants judged whether a sentence was correct or incorrect.  
    \item \textbf{Travel Planning:} Tasks were adapted from the TravelPlanner dataset \cite{Xie2024TravelPlanner}. Our implementation focused on planning a single-day itinerary using structured information from the dataset.  
    \item \textbf{Visual Question Answering (VQA):} Questions were sampled from the OK-VQA dataset \cite{marino2019ok}. From 10 randomly sampled categories, one question was drawn per category, covering a range of everyday visual knowledge queries.  
\end{itemize}

These three tasks were selected to represent ecologically valid AI use cases (language editing, planning, and knowledge retrieval), span distinct cognitive modalities (text, vision, and multi-constraint reasoning), and allow controlled manipulation of AI performance given ground truth labels.

\subsection*{Experimental Setup: Sequential Multi-Task Human–AI Interaction}
Participants completed three consecutive tasks in a randomized order, yielding a fully within-subject multi-task design. AI performance was pre-scripted to fixed levels at the task level: grammar error detection at 30\%, travel planning at 60\%, and VQA at 90\% accuracy. These levels were selected to reflect two principles. First, they capture the relative difficulty of the three domains: grammar error detection remains challenging for LLMs when compared against gold standards, multi-constraint planning tasks typically yield moderate success, and knowledge-oriented VQA tasks are often answered with relatively high accuracy (especially for commonsense queries). This ordering is broadly consistent with benchmark results for CoNLL-2014 (grammar), TravelPlanner, and OK-VQA datasets \cite{ng-etal-2014-conll,Xie2024TravelPlanner,marino2019ok}. Second, the chosen values provide clearly distinguishable performance bands that enable systematic contrasts in belief formation and delegation behavior (choosing to rely on the AI vs. act independently).


We further acknowledge that our operationalization of accuracy levels does not aim to reproduce real-world LLM performance distributions in a literal sense. Real systems exhibit fluctuating and context-dependent accuracy, making it impossible to enforce stable likelihood signals across trials. Our choice of 30/60/90\% therefore reflects stylized yet plausible anchors that preserve the empirical ordering of difficulty across domains while allowing controlled Bayesian benchmarking.

Our primary estimand concerns how beliefs about a multipurpose AI evolve within individuals as they transition across tasks. Because belief updating is path dependent, posteriors formed in one task serve as priors for subsequent tasks thus estimating cross task belief transfer requires observing the same participant under multiple accuracy regimes. A design in which accuracy levels are shuffled across participants, such that each individual experiences only one accuracy per domain, would permit estimation of average accuracy effects but would eliminate the counterfactual needed to test whether beliefs formed in one task systematically influence expectations and delegation in another. The fully within-subject design  enables identification of belief carryover dynamics that cannot be recovered from between-subject or partially crossed designs. 

Task order was fully counterbalanced so that each task (and thus its associated accuracy level) appeared equally often in first, second, and third position.
This design allows us to test whether participants treat the AI consistently across domains or adapt their beliefs based on task-specific experience. Following prior HCI research, these scripted accuracy levels served as controlled anchors balancing ecological plausibility with experimental manipulation \cite{dzindolet2003role,madhavan2007effects}.

\subsection{Procedure}
Upon providing informed consent, participants completed a pre-survey measuring propensity to trust in automation (TiA) \cite{korber2018theoretical}, Need for Cognition (NCS-6) \cite{lins2020very}, and AI literacy (MAILS) \cite{carolus2023mails}. These measures were collected to test whether stable individual differences shape participants’ initial priors about the AI and moderate their subsequent belief updating and delegation behavior. Then the participants were assigned randomly to 1 of 6 order conditions. Before beginning each task, they also had to go through a tutorial of the task to help familiarize themselves with the task and the task interface, addressing potential familiarity biases.

Each task consisted of 10 trials. AI errors were randomly positioned within the block while preserving the block-level accuracy target. 
Further,
\begin{enumerate}[leftmargin=*]
    \item At the beginning of each task (Trial 1), participants reported:
    \begin{itemize}
        \item \textbf{Self-confidence} in solving the task (0–100).  
        \item \textbf{Prior belief} about the AI’s accuracy in the task (0–100).  
        \item Two \textbf{counterfactual priors}: their belief about AI accuracy if its next answer were correct $b_0^+$, and if it were incorrect $b_0^-$(0–100 each).  
        \item Two \textbf{Trust in Automation items} (1–5 Likert): “I trust the system …” and “I can rely on the system …”.  
    \end{itemize}
    \item In each trial, they chose whether to \textbf{delegate} to the AI or answer themselves.  
    \item After submission, participants received feedback showing the AI’s answer, their own answer (if provided), and the correct ground truth. Ground truths were taken directly from benchmark datasets for the Grammar and VQA tasks, while for the Travel Planning task, they were manually constructed to ensure a unique, verifiable itinerary for each item.
    \item Participants then reported their \textbf{posterior belief} about the AI’s accuracy on the next trial (0–100), self-confidence (0-100), and re-rated the two \textbf{Trust in Automation} items after each trial.  
\end{enumerate}

This procedure yielded a dense, trial-by-trial record of confidence ratings, delegation decisions, belief trajectories, and evolving trust. After each task block, participants completed the `Overall Trust' subscale of the Trust in Automation questionnaire. At the end of the study, they filled out a post-experiment questionnaire about their overall experience.

Our design choice to provide participants with the AI’s answer, their own answer, and the correct ground truth was motivated by methodological considerations: it allowed us to isolate and measure belief updating and delegation processes under conditions where accuracy could be objectively verified. Without such ground‐truth feedback, it would be difficult to disentangle whether participants’ trust adjustments were driven by the AI’s performance or by uncertainty about the task itself. Immediate feedback ensures that each trial corresponds to a well-defined likelihood signal, which is necessary for benchmarking human updates against the Beta–Binomial Bayesian model used in our hypotheses. Importantly, many real-world AI applications offer similar opportunities for users to verify correctness—such as code execution, grammar checking, fact-checking with citations, or validating structured outputs—meaning that providing outcome feedback enhances diagnostic clarity without altering the underlying structure of the interaction.

\begin{figure*}[!h]
\centering
\begin{minipage}{0.32\textwidth}
    \centering
    \includegraphics[width=\linewidth]{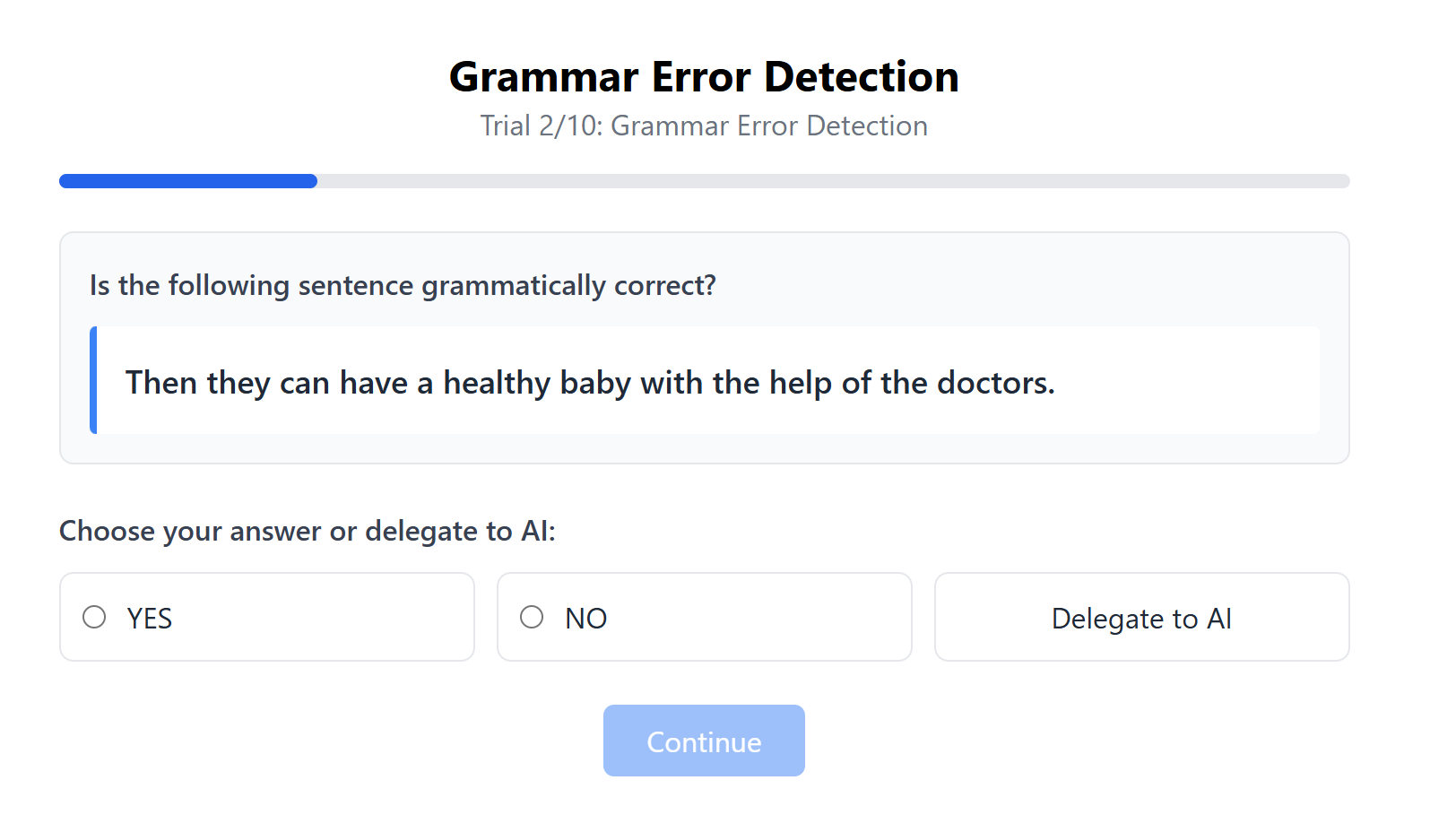}
    \caption*{(a) Grammar Error Detection}
\end{minipage}\hfill
\begin{minipage}{0.32\textwidth}
    \centering
    \includegraphics[width=\linewidth]{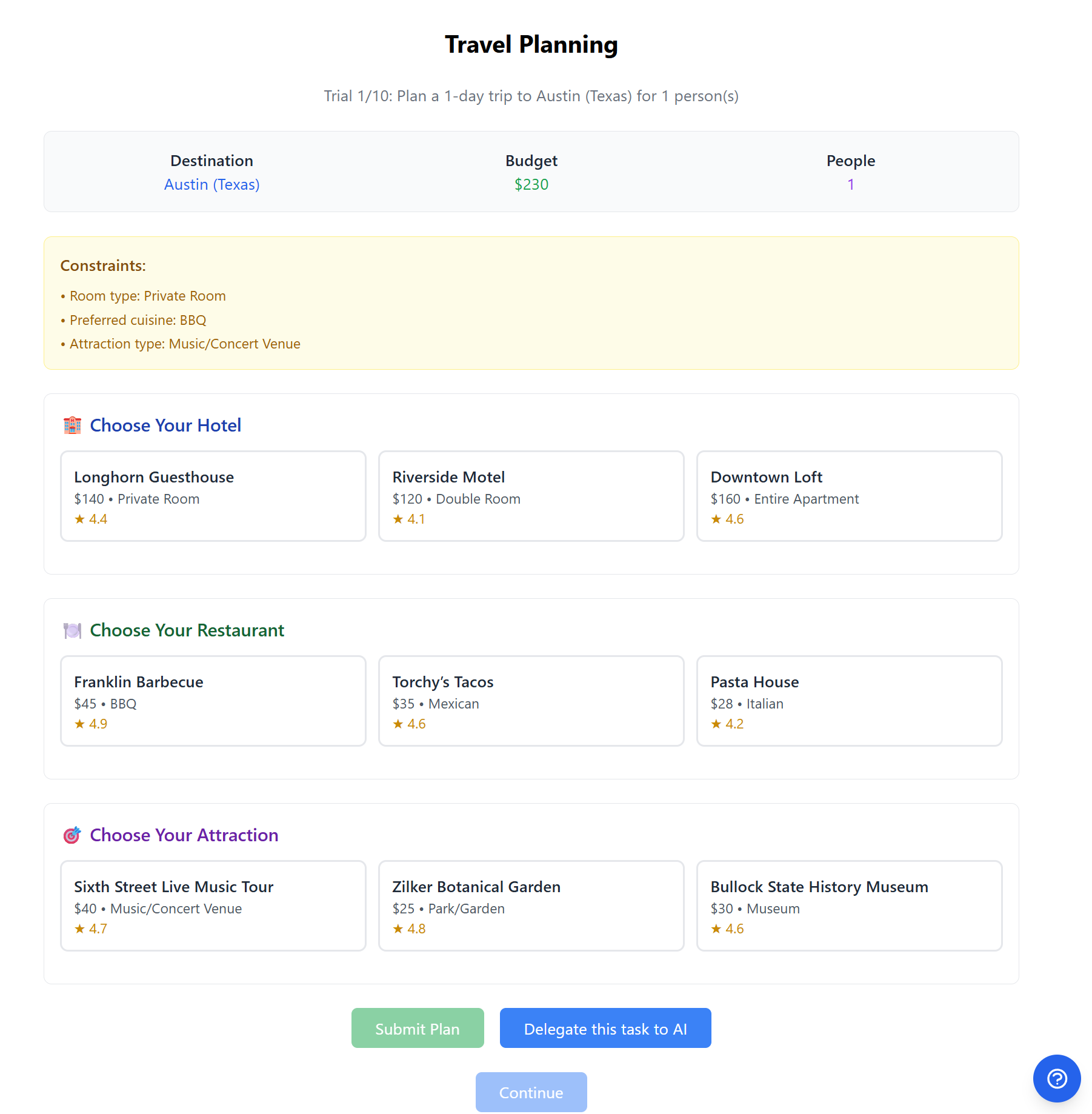}
    \caption*{(b) Travel Planning}
\end{minipage}\hfill
\begin{minipage}{0.32\textwidth}
    \centering
    \includegraphics[width=\linewidth]{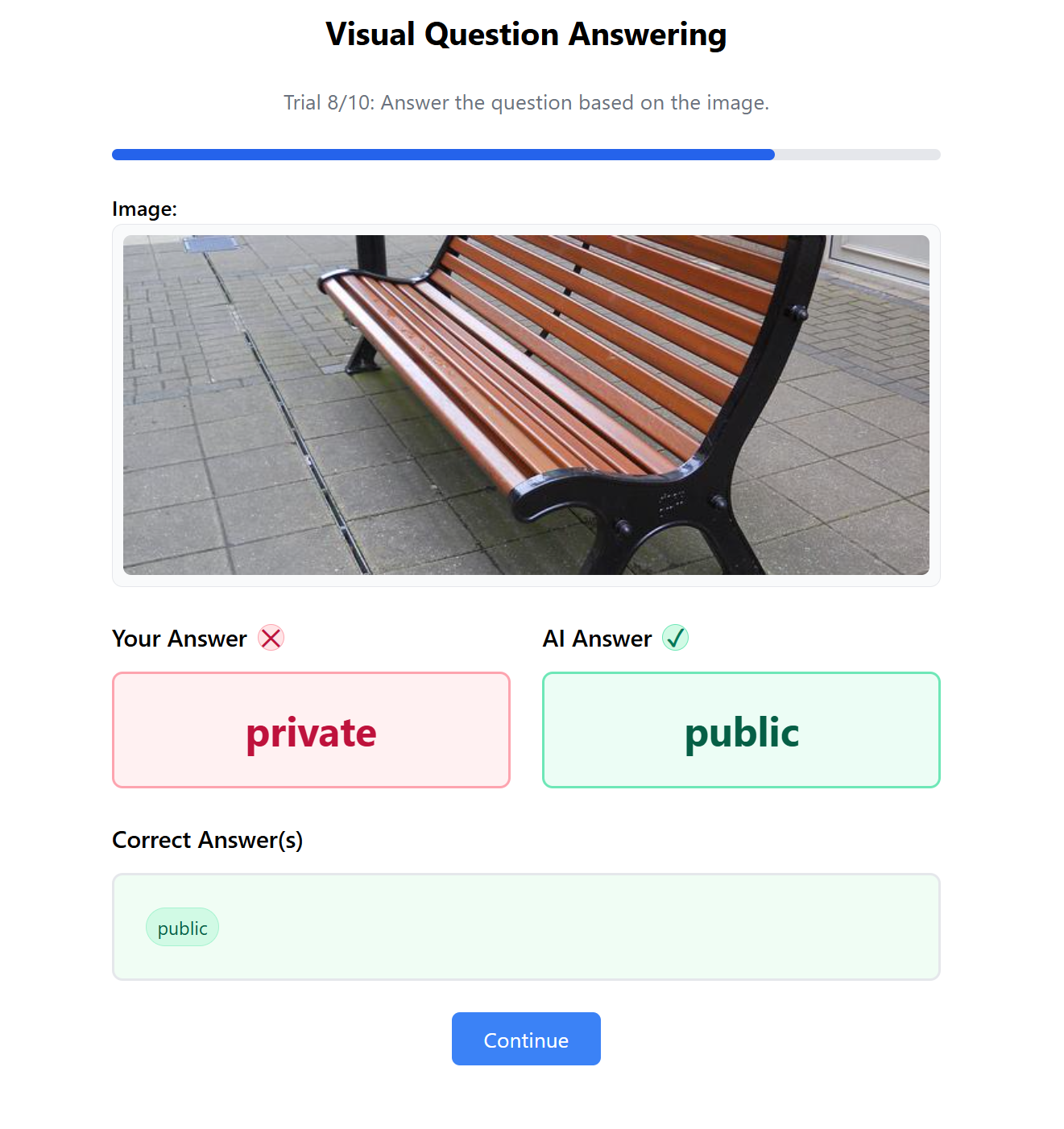}
    \caption*{(c) Visual Question Answering}
\end{minipage}
\caption{Screenshots of the three experimental tasks. Participants completed (a) Grammar error detection, (b) travel planning, and (c) visual question answering tasks in a randomized order.}
\label{fig:task_screens}
\end{figure*}

\subsection{Participants}
We recruited 240 participants via Prolific, restricted to U.S. adults with at least 100 prior submissions and a 90\% approval rating. Participants received a base payment of £2.25 and up to £3 in performance-based bonus, ensuring fair hourly compensation of ~£8–12 (including bonuses)~\cite{kaur2026incentive}. Attention checks and comprehension questions were included; participants who failed them were screened out. In total, we collected 293 submissions. Of these, 18 participants failed comprehension checks and 35 violated integrity requirements (e.g., engaged in prohibited attempts to access backend solutions), leaving 240 valid participants for the final analysis.

The final sample ($N=240$) had a mean age of $M = 42.31$ ($SD = 13.42$). Of these, 120 identified as female, 119 as male, and 1 as non-binary. Educational attainment included: 41.67\% bachelor’s degree, 29.17\% high school, 20.42\% master’s degree, 4.17\% doctorate, and 4.58\% other/prefer not to disclose.

\subsection{Main Measures}
We analyze participants’ interactions using the following primary outcomes:  

\noindent\emph{\textbf{Belief Updating}.}  
Trial-by-trial priors and posteriors on AI accuracy (0–100) test for Bayesian vs. bounded-rational updating (H2).  

\noindent\emph{\textbf{Delegation}.}  
A binary choice each trial (self vs. AI) serves as the primary behavioral measure of reliance. Delegation models test how lagged beliefs and confidence predict reliance (H3, H4).  

\noindent\emph{\textbf{Confidence}.}  
Self-reported confidence (0–100) provides insight into self-assessment and how it interacts with reliance.  

\noindent\emph{\textbf{Trust and Dispositional Measures}.}  
TiA (pre/post trust, reliance), NCS-6, and MAILS measure stable individual differences. We test whether dispositional trust predicts higher initial priors (H5).

\begin{table*}[!ht]
\centering
\caption{Summary of variables collected across the experiment.}
\label{tab:variables}
\renewcommand{\arraystretch}{1.15}
\begin{tabularx}{\linewidth}{p{3.2cm} p{3.2cm} >{\raggedright\arraybackslash}X}
\toprule
\textbf{Variable (notation)} & \textbf{Type/Scale} & \textbf{Question Shown to Participants} \\
\midrule
\textit{Self-confidence (prior)} (\(conf_{\text{pre}}\)) & 0--100 slider/number & ``How confident are you in your ability to complete this task correctly?'' \\
\textit{AI prior belief} (\(b_{0}\)) & 0--100 slider/number & ``How likely do you think it is that the AI will answer correctly in the next trial?'' \\
\textit{Counterfactual prior (correct)} (\(b_{0}^{+}\)) & 0--100 slider/number & ``Imagine the AI's \textbf{next} answer is \textbf{correct}. After that, what would you believe is the AI's chance of being correct on the \textbf{following trial}?'' \\
\textit{Counterfactual prior (incorrect)} (\(b_{0}^{-}\)) & 0--100 slider/number & ``Imagine the AI's \textbf{next} answer is \textbf{incorrect}. After that, what would you believe is the AI's chance of being correct on the \textbf{following trial}?'' \\
\textit{Trust in Automation (pre)} (\(trust_{\text{pre}}\)) & 1--5 Likert & ``I \textbf{trust} the system that will assist me in completing this task.'' (1 = strongly disagree, 5 = strongly agree) \\
\textit{Reliance in Automation (pre)} (\(rely_{\text{pre}}\)) & 1--5 Likert & ``I can \textbf{rely} on the system that will assist me in completing this task.'' (1 = strongly disagree, 5 = strongly agree) \\
\textit{Self-confidence (posterior)} (\(conf_{\text{post}}\)) & 0--100 slider/number & ``How confident are you in your ability to complete this task correctly?'' (asked after each trial) \\
\textit{AI posterior belief} (\(b_{\text{post}}\)) & 0--100 slider/number & ``How likely do you think AI is to answer correctly on the next trial?'' \\
\textit{Trust in Automation (post)} (\(trust_{\text{post}}\)) & 1--5 Likert & ``I \textbf{trust} the system that will assist me in completing this task.'' (1 = strongly disagree, 5 = strongly agree) \\
\textit{Reliance in Automation (post)} (\(rely_{\text{post}}\)) & 1--5 Likert & ``I can \textbf{rely} on the system that will assist me in completing this task.'' (1 = strongly disagree, 5 = strongly agree) \\
\textit{Delegation decision} (\(delegate\)) & Binary (0=self, 1=AI) & ``Do you want to answer this yourself, or let the AI answer for you?'' \\
\bottomrule
\end{tabularx}
\end{table*}

%% file: tex/4_Analysis.tex
\section{Statistical Modeling and Analysis}

Our analysis was preregistered on OSF\footnote{\url{https://osf.io/7b84a/?view_only=95f6fdacaca2414c8c3d0b6328059212}}. All de-identified data, analysis code, and study materials will be made publicly available in the same OSF repository upon publication. We tested five hypotheses, corresponding to the research questions outlined earlier. All models were estimated with standard errors clustered at the participant level to account for repeated measures. Unless otherwise noted, robustness checks included OLS, GEE, and mixed-effects regressions. For all analyses, only valid participants who completed the full trial and cleared attention checks and task-related comprehension questions were considered.

\subsection{H1: Cross-Task Priors}

\paragraph{Data preparation.}  
H1 tests whether participants’ prior beliefs about AI accuracy in a new task ($b0_{\text{curr}}$) depend on their posterior beliefs from the previous task ($bT_{\text{prev}}$). Because this requires participants to have completed at least one task before, we restricted the dataset to tasks in positions 2 or 3. This yielded 480 observations (240 participants × 2 subsequent tasks each).  Belief variables were kept on the 0--100 scale for interpretability. 

\paragraph{Primary model.}  
For our first hypothesis we estimated a mixed-effects regression predicting the initial prior in each task $k$ from the belief at the end of the preceding task:
\[
b0_{\text{curr}} \sim bT_{\text{prev}} + C(\text{taskType}) + C(\text{taskPosition}) + (1 \mid \text{participant}),
\]
with random intercepts at the participant level to capture individual heterogeneity. A rational Bayesian benchmark predicts $\beta \approx 0$, indicating full resetting of priors across unrelated tasks. A positive $\beta$ indicates anchoring, while a negative $\beta$ would imply overcompensation. As a robustness check, we also estimated pooled OLS with participant-clustered standard errors.

\paragraph{Additional analyses.}  
To test robustness, we implemented three complementary specifications:
\begin{itemize}
    \item \textbf{Objective performance model:} Replacing $bT_{\text{prev}}$ with the actual accuracy of the previous task (proportion of correct trials). This tests whether carryover reflects participants’ subjective beliefs or the true performance they observed.  
    \item \textbf{Belief-change model:} Modeling the difference $b0_{\text{curr}} - bT_{\text{prev}}$ as outcome, predicted by task type and position. This isolates whether participants systematically reset upwards or downwards relative to their previous posterior.  
    \item \textbf{Individual differences:} Adding pre-survey covariates (MAILS for AI literacy, NCS-6 for need for cognition) to ensure that cross-task carryover is not confounded by stable participant traits.  
\end{itemize}

This triangulation allows us to distinguish whether priors in a new task reflect rational resetting, objective performance learning, or boundedly rational anchoring to prior experiences.

\subsection{H2: Bayesian vs. Bounded-Rational Updating}

\paragraph{Data preparation.}
To test H2, we examine whether belief updates within each task follow Bayesian rationality. For each participant–task, we constructed trial-by-trial belief changes 
\[
\Delta b_t = b^{\text{post}}_t - b^{\text{pre}}_t,
\]
and computed the corresponding Bayesian-normative update step
\[
\Delta b^{\text{Bayes}}_t = \frac{f_t - m^B_{t-1}}{n_0 + t},
\]
where $f_t \in \{0,1\}$ indicates the AI outcome (correct/incorrect), $m^B_{t-1}$ is the Bayesian mean belief prior to trial $t$, and $n_0$ is the prior precision (stickiness), estimated from counterfactual priors ($b_0^+, b_0^-$) that reveal each participant’s expected shift after a single correct vs.\ incorrect outcome. \textbf{Concretely, we infer a Beta prior $(\alpha,\beta)$ such that $b_0=\alpha/(\alpha+\beta)$ and the one-step posteriors after a correct/incorrect outcome match $(b_0^+, b_0^-)$; this yields $n_0=\alpha+\beta$.} (Here, $t$ counts observed outcomes starting at $1$, so the one-step Bayesian update scales by $1/(n_0 + t)$.) Beliefs were modeled on their original 0--100 scale for interpretability, so coefficients reflect percentage-point changes rather than unit-normalized values. Because counterfactual priors can fail quality checks (out-of-range, monotonicity violation, degeneracy), we implemented three preregistered QC policies—\emph{strict}, \emph{lenient}, and \emph{hybrid} with fallback priors—to derive $n_0$ and test robustness.

\paragraph{Primary model.}  
We estimate the regression
\[
\Delta b_t \sim \sigma \cdot \Delta b^{\text{Bayes}}_t,
\]
demeaning both $\Delta b_t$ and $\Delta b^{\text{Bayes}}_t$ within each participant–task to remove baseline differences. If participants update rationally, $\sigma$ should equal 1; systematic deviations indicate bounded rationality (conservatism if $\sigma<1$, overreaction if $\sigma>1$). We fit pooled OLS with participant-clustered SEs and, as a robustness check, participant-level mixed models.

\paragraph{Additional analyses.}  
We conducted several complementary tests:
\begin{itemize}[leftmargin=*]
    \item \textbf{Per-task slopes:} Estimating $\sigma$ separately for grammar (30\%), travel (60\%), and VQA (90\%) tasks to assess domain-specific updating.  
    \item \textbf{Individual slopes:} Extracting per-participant updating coefficients to characterize heterogeneity.  
    \item \textbf{Robustness across CF policies:} Repeating the analysis under strict, lenient, and hybrid counterfactual policies with fallback priors.  
    \item \textbf{Visualization:} Histograms of individual slopes and scatter plots of observed vs. normative steps, with Bayesian $\sigma=1$ as reference.  
\end{itemize}

\subsection{H3: Beliefs and Delegation Decisions} 

\paragraph{Data preparation.} 
For H3, we test whether participants’ delegation choices (self vs.\ AI) are predicted by their lagged beliefs about AI accuracy. The outcome variable is binary ($delegate=1$ if the AI was chosen). To ensure proper temporal ordering, we lagged beliefs by one trial: the belief reported before trial $t$ predicts delegation on trial $t$. For the first trial of each task, we use the pre-task prior belief as the lag. Beliefs were normalized to $[0,1]$ for estimation; effects are reported and plotted in percentage points for interpretability. 

\paragraph{Primary model.} 
We estimated logistic regressions with participant-clustered standard errors of the form:
\[
delegate_{it} \sim b_{i,t-1} + C(\text{taskType}) + C(\text{trialOrdinal}) + C(\text{taskPosition}),
\]
where $b_{i,t-1}$ denotes the participant’s lagged belief about AI accuracy. Task fixed effects capture domain differences, trial fixed effects control for within-task learning, and task-order fixed effects control for between-task order. As a robustness check, we also estimated generalized estimating equations (GEE) with exchangeable correlation at the participant level. A rational benchmark predicts that delegation increases monotonically with belief in AI accuracy.

\paragraph{Alternative specifications.} 
To ensure robustness, we tested: 
\begin{itemize} 
    \item \textbf{Task-specific regressions:} Estimating models separately for grammar, travel, and VQA to assess whether belief–delegation links differ by domain.  
    \item \textbf{Objective performance:} Adding the AI’s actual correctness on the previous trial as a predictor, to test whether delegation is driven by subjective beliefs or by objective accuracy signals.  
    \item \textbf{Individual differences:} Adding pre-survey measures (TiA, MAILS, NCS-6) to test whether dispositional trust or cognitive style moderate belief–delegation sensitivity.  
\end{itemize} 

Together, these specifications triangulate whether delegation is primarily belief-driven, domain-specific, or shaped by stable individual traits.

\subsection{H4: Confidence and Delegation Decisions}

\paragraph{Data preparation.}  
For H4, we test whether participants’ self-reported confidence influences their delegation behavior. The outcome variable is binary ($delegate{=}1$ if the AI was chosen). To respect temporal ordering, we lagged self-confidence by one trial: $conf_{t-1}$ predicts delegation at trial $t$. For the first trial of each task, we used participants’ \emph{pre-task confidence} as the lagged value. Confidence was normalized to $[0,1]$ for estimation; effects are reported and plotted in percentage points for interpretability. 

\paragraph{Primary model.}  
We estimated logistic regressions with participant-clustered standard errors of the form:
\[
delegate_{it} \sim conf_{i,t-1} + C(\text{taskType}) + C(\text{trialOrdinal}) + C(\text{taskPosition}),
\]
where $conf_{i,t-1}$ denotes the lagged self-confidence. Task fixed effects capture domain differences, trial fixed effects control for within-task learning, and task-order fixed effects control for between-task order. As a robustness check, we also estimated generalized estimating equations (GEE) with exchangeable correlation at the participant level. A rational benchmark predicts that delegation should decrease with higher self-confidence.

\paragraph{Alternative specifications.}  
To probe robustness, we estimate three variants:
\begin{itemize}
    \item \textbf{Joint model:} Including both lagged confidence and lagged beliefs simultaneously, to test whether confidence exerts an independent effect once beliefs are controlled.  
    \item \textbf{Task-specific models:} Running separate regressions for grammar, travel, and VQA to examine whether the role of confidence varies by domain.  
    \item \textbf{Individual differences:} Adding TiA, MAILS, and NCS-6 scores as moderators to assess whether dispositional trust or cognitive style conditions the confidence–delegation link.  
\end{itemize}

These specifications help determine whether confidence is a direct predictor of delegation, a proxy for beliefs, or a domain-dependent moderator of reliance behavior.

\subsection{H5: Dispositional Trust and Initial Priors}

\paragraph{Data preparation.}  
To test whether dispositional trust predicts higher initial beliefs about AI accuracy, we constructed a participant–task panel using trial-1 priors from each of the three tasks. Each row corresponds to a participant $\times$ task, yielding 720 observations (240 participants $\times$ 3 tasks). These priors (0--100 scale) were merged with pre-survey measures: propensity to trust in automation (TiA), Need for Cognition (NCS-6), and AI literacy (MAILS). For robustness, we also estimated standardized models; all main coefficients are reported in percentage-points for interpretability.  

\paragraph{Primary model.}  
We fit a linear mixed-effects regression with participant random intercepts:  
\[
\text{prior}_{it} \sim \text{TiA}_{i} + C(\text{taskType}) + C(\text{taskPosition}) + u_i,
\]  
where $u_i$ captures participant-level heterogeneity. This specification tests whether individuals with higher TiA scores report systematically higher priors, while accounting for task differences and order effects.  

\paragraph{Robustness models.}  
Two complementary specifications provide checks:  
\begin{enumerate}
    \item \textbf{Control model:} Adds MAILS and NCS-6 as covariates to test whether TiA effects persist net of broader cognitive style and AI literacy.  
    \item \textbf{OLS fallback:} Ordinary least squares with participant-clustered SEs, to guard against occasional convergence issues in MixedLM.  
\end{enumerate}

\paragraph{Summary.}  
This setup isolates the role of dispositional trust in shaping accuracy priors before any feedback is observed. If rational agents set priors independently of personality traits, TiA should not matter. A positive TiA coefficient, by contrast, indicates that dispositional trust colors baseline expectations about AI performance---consistent with H5.

%% file: tex/5_Results.tex
\section{Results}

We analyze participants’ priors, belief updating, and delegation behaviors across 240 participants (7200 trials, 3 tasks per participant). For each hypothesis, we present the statistical models, robustness checks, and main findings.

\subsection{H1: Do priors reset across tasks?}

We first tested whether participants reset their prior beliefs about AI accuracy when starting a new task, or if priors were anchored to their most recent posterior belief from the previous task. 

\begin{figure}[t]
  \centering

  \begin{subfigure}[b]{\linewidth}
    \centering
    \includegraphics[width=\linewidth]{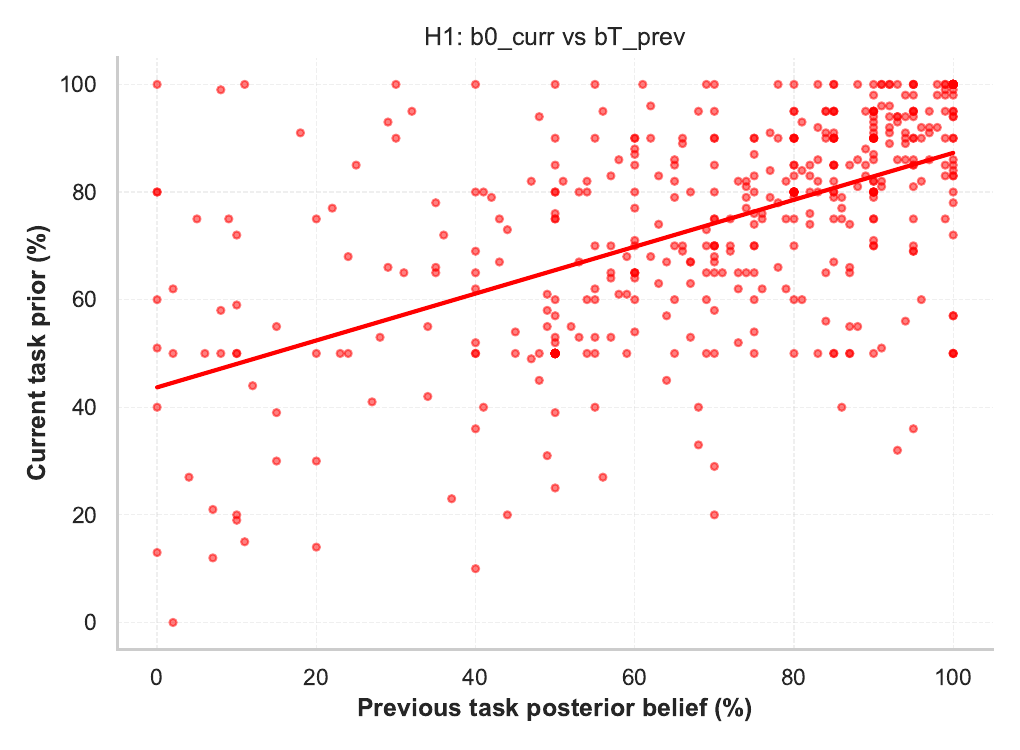}
    \subcaption{Cross-task carryover of beliefs (H1). Each point is one participant--task observation (positions 2--3 only; $N=480$). The line is an OLS fit for visualization (unclustered). Slope $\approx 0.43$; in models with SEs clustered by participant the coefficient on previous posterior is $\beta=0.425$, $SE=0.046$, $p< .001$ (Table~\ref{app:h1_ols_robust}).}
    \label{fig:h1_scatter}
  \end{subfigure}
  \hfill
  \begin{subfigure}[b]{\linewidth}
    \centering
    \includegraphics[width=\linewidth]{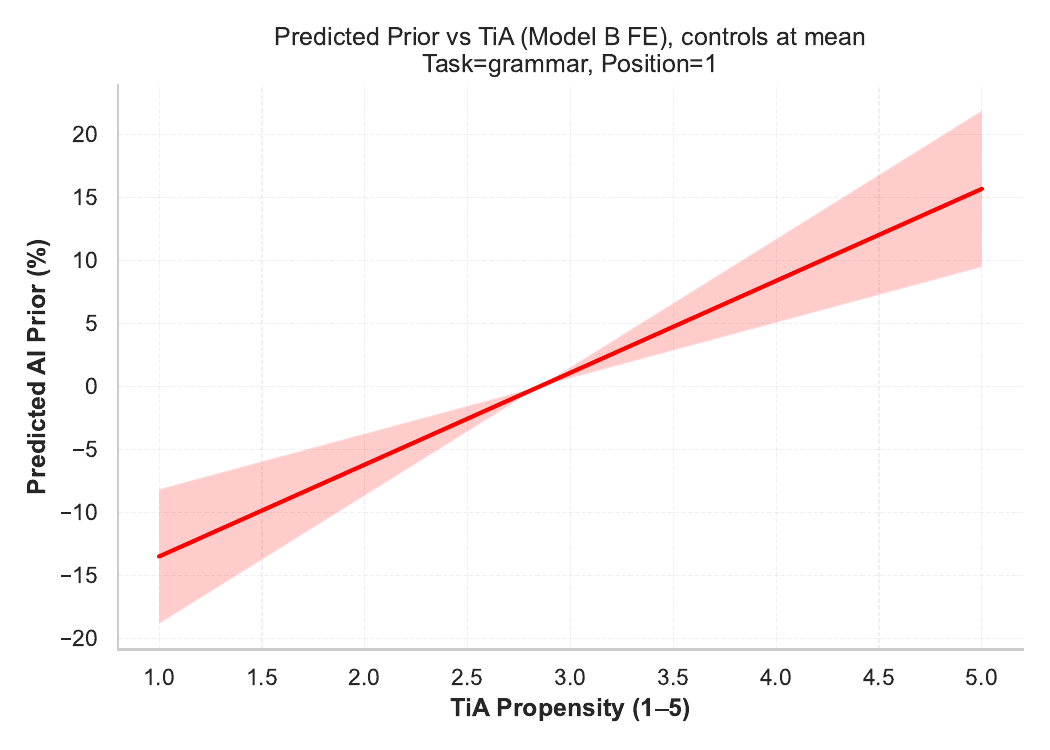}
    \subcaption{Effect of dispositional trust (TiA) on prior beliefs about AI accuracy. Each point on the x-axis is a one-unit increase on the 1--5 TiA scale; the y-axis shows predicted AI priors on the 0--100 scale. Shaded area indicates 95\% CI.}
    \label{fig:h5_tia_marginal}
  \end{subfigure}

\end{figure}

\paragraph{Main effect.}  
As preregistered, we estimated a mixed-effects model with participant random intercepts. The analysis reveals strong cross-task carryover: a 10-point higher posterior in the previous task predicts a 3.0-point higher prior in the subsequent task ($\beta = 0.296$, $SE = 0.043$, $p < .001$; Table~\ref{tab:h1_mixed}). Thus, instead of resetting, participants imported expectations across tasks. For robustness, we also estimated a pooled OLS regression with participant-clustered standard errors, which yielded a highly consistent pattern with a somewhat larger coefficient ($\beta = 0.432$, $SE = 0.045$, $p < .001$; Appendix~\ref{app:h1_ols}). Figure~\ref{fig:h1_scatter} visualizes this spillover: priors in task $t{+}1$ rise systematically with the posterior from task $t$.

\paragraph{Robustness.}  
To test the robustness of this effect, we implemented three complementary checks.  
\begin{itemize}[leftmargin=*]
    \item \textbf{Objective performance.} Substituting the previous task’s \emph{objective} AI accuracy instead of participants’ posterior beliefs produced no significant effect ($\beta = 0.018$, $SE = 0.024$, $p = .459$; Appendix~\ref{app:h1_ols_robust}), indicating that priors are shaped by subjective impressions rather than the ground-truth accuracy. Notably, when subjective posteriors were excluded from the model, objective accuracy did predict subsequent priors (Appendix~\ref{app:h1_mixed_prevObj}); this pattern suggests that participants only integrate objective performance insofar as it shapes their subjective posterior.

    \item \textbf{Belief change.} Modeling the difference $b0_{\text{curr}} - bT_{\text{prev}}$ showed that priors were on average slightly lower than the preceding posterior (intercept $-2.56$, $p = .204$). This reset, however, was incomplete: participants made significant upward adjustments in travel ($\beta = 7.36$, $p = .003$) and VQA ($\beta = 9.03$, $p < .001$) relative to grammar (Appendix~\ref{app:h1_change}).

    \item \textbf{Individual heterogeneity.} Adding pre-survey controls (MAILS, NCS-6) left the coefficient on $bT_{\text{prev}}$ virtually unchanged, confirming that cross-task carryover is robust to participant-level differences (Appendix~\ref{app:h1_controls}).

\end{itemize}

\begin{framed}
\paragraph{Summary.}  
Across specifications, priors in a new task were strongly predicted by participants’ most recent posterior beliefs. A one–standard-deviation increase in the previous posterior carried over to roughly a one–third–standard-deviation increase in the next prior, whereas objective AI performance exerted no independent influence once beliefs were accounted for. This pattern violates the preregistered rational benchmark of full resetting (\(\beta = 0\)) and demonstrates systematic \emph{belief inertia}: participants anchor on past impressions even when facing unrelated tasks with different true accuracy levels. Rather than rationally ignoring irrelevant history, participants imported expectations across tasks, producing spillover that reflects bounded updating rather than Bayesian rationality.
\end{framed}

\begin{table}[H]
\centering
\caption{Appendix H1. Mixed-effects regression of current-task priors ($b0_{\text{curr}}$).}
\label{tab:h1_mixed}
\begin{tabular}{lcc}
\toprule
 & Coef. & SE \\
\midrule
Previous posterior ($bT_{\text{prev}}$) & 0.296*** & (0.043) \\
Task FE & Yes &  \\
Position FE & Yes &  \\
Random intercept (participant) & Yes & \\
\midrule
Observations & 480 & \\
Log-likelihood & -2026.5 & \\
\bottomrule
\multicolumn{2}{l}{\footnotesize Dependent variable is $b0_{\text{curr}}$.}\\
\multicolumn{2}{l}{\footnotesize *** $p<0.001$.}\\
\end{tabular}
\end{table}


\subsection{H2: Bayesian Updating Within Tasks}

\paragraph{Main effect.}  
As preregistered, we tested whether belief updates within each task followed Bayesian rationality. Under the \emph{strict} counterfactual policy, participants updated in the correct direction but at only half the normative rate: the pooled regression yielded $\hat{\sigma}=0.52$ ($SE=0.06$, $p<.001$, $N=3{,}430$ updates; Table~\ref{tab:h2_results}). This slope below one indicates systematic under-reaction relative to Bayes’ rule.

\paragraph{Robustness.}  
Results were highly consistent across alternative counterfactual policies. With the \emph{lenient} policy, the updating coefficient was $\hat{\sigma}=0.47$ ($SE=0.05$, $p<.001$, $N=5{,}670$). Under the \emph{hybrid} policy with fallback priors, estimates remained in the same range: $\hat{\sigma}=0.47$ ($SE=0.04$) for $S=5$, $\hat{\sigma}=0.49$ ($SE=0.05$) for $S=10$, and $\hat{\sigma}=0.49$ ($SE=0.05$) for $S=20$ (all $p<.001$, $N=6{,}560$; Appendix~\ref{app:h2_robust}). Per-task regressions (grammar, travel, VQA) and distributions of individual-level slopes further confirmed this pattern (Appendix~\ref{app:h2_taskwise}).
\begin{framed}

\paragraph{Summary.}  
Across specifications, participants exhibited systematic \textit{conservatism bias}: they updated in the normative direction, but only about 50\% of the Bayesian step. This under-reaction echoes classic findings in belief-updating research \cite{Edwards1968,kahneman1972subjective}, here replicated in a multi-task Human–AI setting. While participants responded to AI successes and failures, they integrated feedback cautiously rather than fully, leading to slower convergence to the AI’s true accuracy.
\end{framed}

\begin{table}[!ht]
\centering
\caption{Regression of observed belief updates on Bayesian normative steps (H2). Clustered SEs at participant level.}
\label{tab:h2_results}
\begin{tabular}{lcccc}
\toprule
Policy & $N$ updates & $\hat{\sigma}$ & SE & $p$-value \\
\midrule
Strict (CF only)       & 3,430 & 0.52 & 0.06 & $<.001$ \\
Lenient (CF only)      & 5,670 & 0.47 & 0.05 & $<.001$ \\
Hybrid (fallback $S=5$)  & 6,560 & 0.47 & 0.04 & $<.001$ \\
Hybrid (fallback $S=10$) & 6,560 & 0.49 & 0.05 & $<.001$ \\
Hybrid (fallback $S=20$) & 6,560 & 0.49 & 0.05 & $<.001$ \\
\bottomrule
\end{tabular}
\end{table}

\begin{figure}[t]
  \centering
  \includegraphics[width=\linewidth]{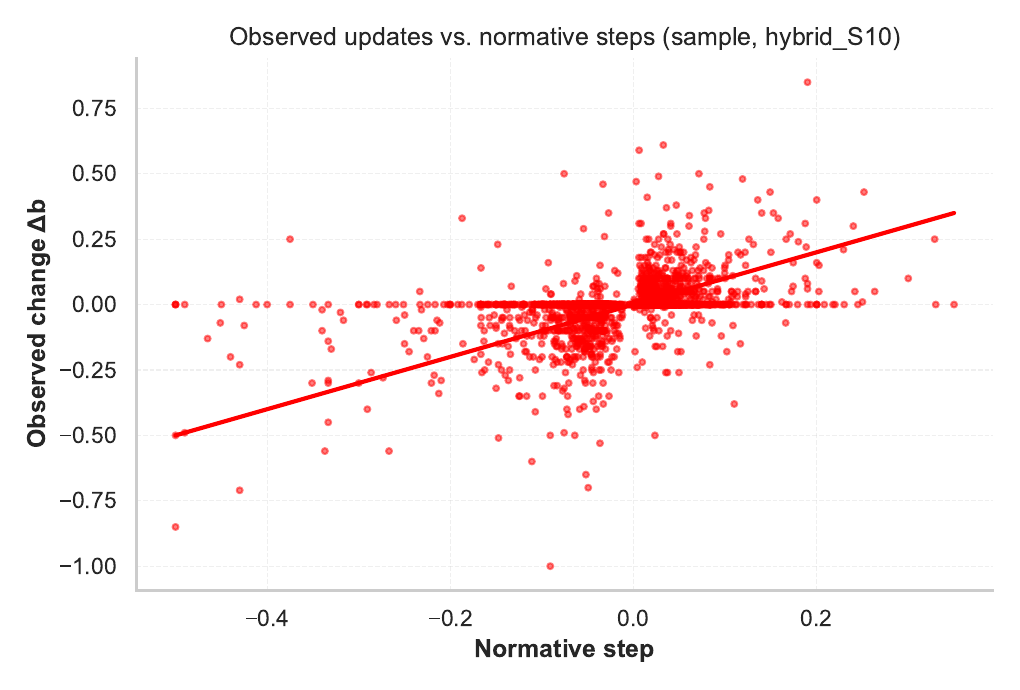}
  \caption{Observed vs.\ normative belief updates under the strict CF policy (H2). 
Each point is one trial-level update (demeaned within participant–task). 
The slope ($\hat{\sigma}\approx 0.52$) indicates conservatism: participants 
updated in the correct direction, but only about half as much as Bayes’ rule prescribes. 
Appendix~\ref{app:h2_robust} shows analogous plots for lenient and hybrid policies.}

  \label{fig:h2_scatter}
\end{figure}

\subsection{H3: Do beliefs influence delegation?}

We next examined whether participants’ beliefs about the AI’s accuracy predicted their delegation decisions.  

\paragraph{Main effect.}  
Across all analyses, stronger beliefs in the AI’s accuracy consistently increased the likelihood of delegation. Participants who believed more strongly that the AI would be correct were substantially more likely to hand over decisions to it ($\beta \approx 1.6$–$1.8$, $SE \approx 0.3$, $p < .001$; Appendix~\ref{app:h3_base}).  

\paragraph{Task differences.}  
Delegation rates were generally higher in the travel and VQA tasks compared to grammar. The influence of belief was strongest in grammar, somewhat reduced in travel, and significantly weaker in VQA, consistent with a ceiling effect in perceptual tasks where delegation was already high. Detailed interaction results are reported in Appendix~\ref{app:h3_interaction}.  

\paragraph{Controls.}  
Adding individual differences did not alter the core findings. General trust in automation predicted greater delegation, whereas AI literacy and need for cognition showed no reliable effects (Appendix~\ref{app:h3_controls}).  
\begin{framed}

\paragraph{Summary.}  
Together, these results demonstrate that participants with stronger beliefs in the AI’s accuracy were more likely to delegate, ceteris paribus. The effect is robust and substantial across models, though attenuated in contexts where delegation is already high.  
\end{framed}

\begin{figure}[t]
  \centering

  \begin{subfigure}[b]{\linewidth}
    \centering
    \includegraphics[width=\linewidth]{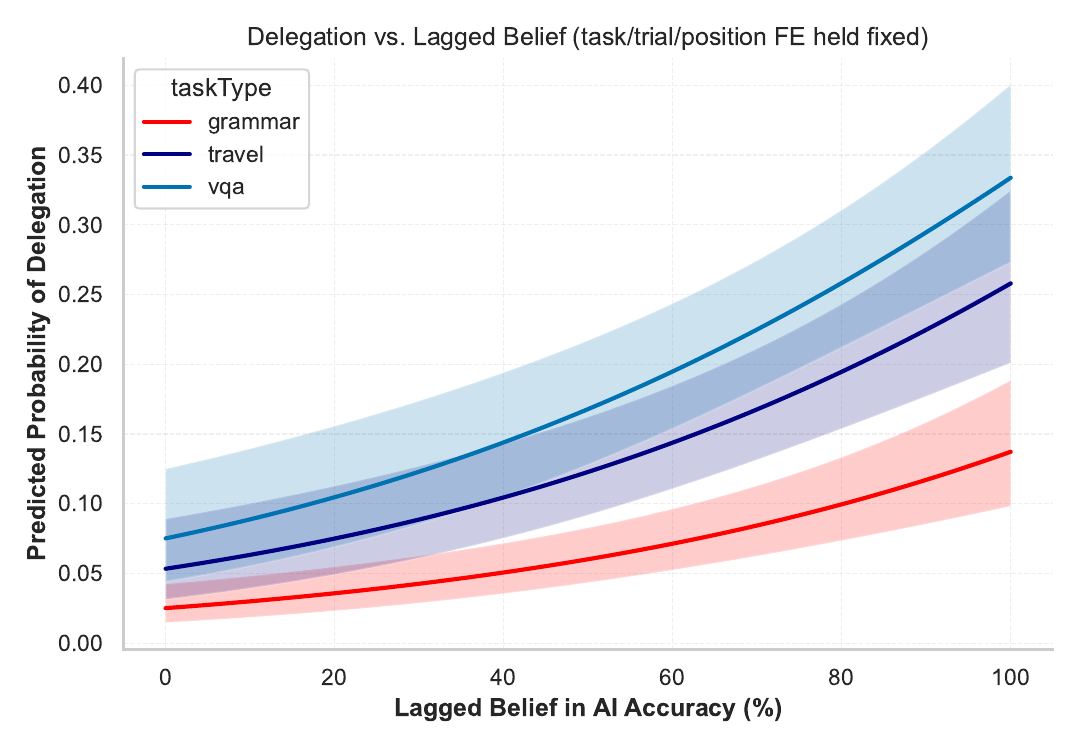}
    \subcaption{Predicted probability of delegation as a function of lagged belief in AI accuracy (H3). Shaded areas show 95\% CIs. Belief increases delegation across all tasks, with weaker effects in VQA.}
    \label{fig:h3_marginal}
  \end{subfigure}
  \hfill
  \begin{subfigure}[b]{\linewidth}
    \centering
    \includegraphics[width=\linewidth]{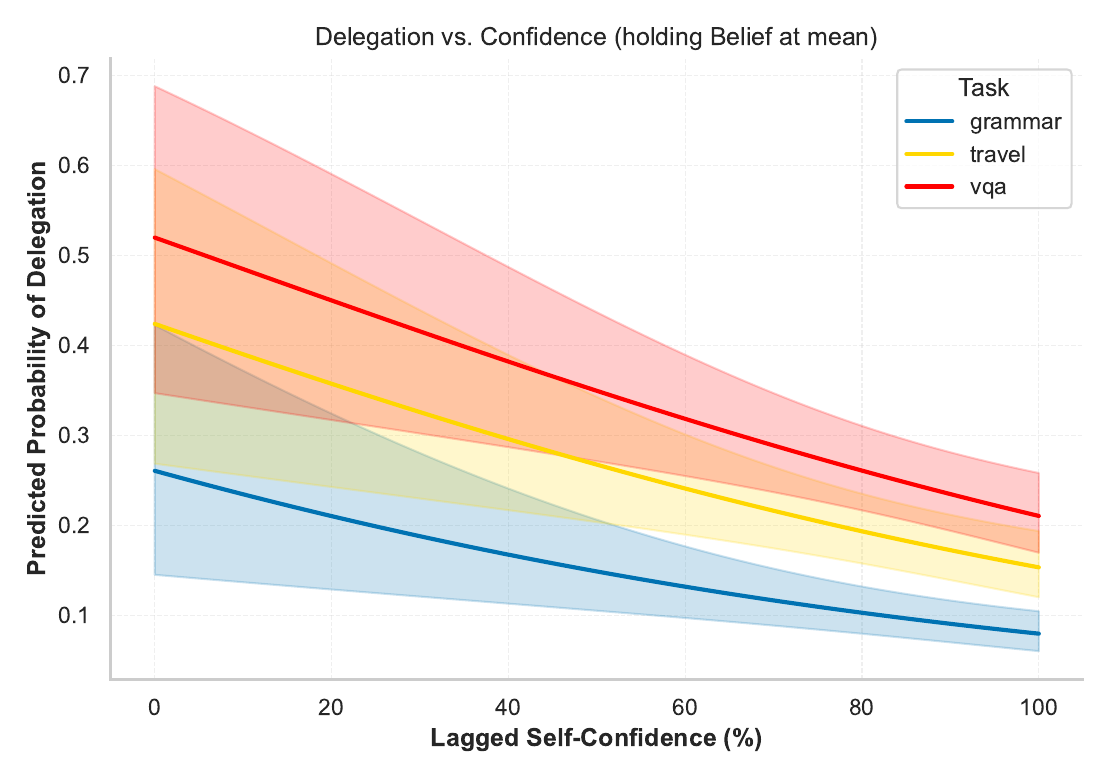}
    \subcaption{Predicted probability of delegation as a function of lagged self-confidence (H4), holding beliefs constant at their mean. Shaded areas show 95\% CIs. Delegation decreases with confidence across all tasks, but the effect is smaller than that of belief (see H3).}
    \label{fig:h4_marginal}
  \end{subfigure}

\end{figure}

\subsection{H4: Does confidence influence delegation?}

\paragraph{Main effect.}  
Lagged self-confidence strongly predicted reduced delegation. In the pooled GEE model, higher confidence significantly decreased the likelihood of delegating to the AI ($\beta=-1.08$, $SE=0.32$, $p=.001$; Appendix~\ref{app:h4_confonly}). In other words, more confident participants were more inclined to keep control rather than rely on the AI. Before trial~1, where we capture pre-task confidence, the effect was directionally negative but not significant ($\beta=-0.23$, $SE=0.62$, $p=.710$; Appendix~\ref{app:h4_preonly}).

\paragraph{Joint model.}  
Including both lagged confidence and lagged beliefs confirmed that confidence exerts an independent effect. Confidence negatively predicted delegation ($\beta=-1.54$, $SE=0.34$, $p<.001$) even after controlling for belief, while belief remained a strong positive predictor ($\beta=1.92$, $SE=0.32$, $p<.001$; Appendix~\ref{app:h4_joint}).
\begin{framed}
\paragraph{Summary.}  
Together, these results indicate that confidence and belief jointly shape delegation. Participants were more likely to delegate when they believed in the AI’s accuracy, but less likely when they felt confident in their own judgment. This supports the preregistered prediction that self-confidence and trust in the AI act as competing drivers of delegation.

\end{framed}


\subsection{H5: Do individual differences in trust propensity predict priors?}

\paragraph{Main effect.}  
We tested whether participants with a higher dispositional propensity to trust automation (TiA) reported higher priors about AI accuracy before any feedback. Mixed-effects models with participant random intercepts confirm a strong positive effect: in Model~A, each one-point increase in TiA (1–5 scale) predicts an 8.43-point higher accuracy prior on the 0–100 scale ($\beta = 8.43$, $SE = 1.44$, $p < .001$; Table~\ref{tab:h5_main}). This effect holds across task types and positions, suggesting that dispositional trust systematically shifts initial expectations (Figure~\ref{fig:h5_tia_marginal}).

\paragraph{Robustness.}
When controlling for additional pre-survey measures (MAILS for AI literacy and NCS-6 for need for cognition), the TiA effect remains strong ($\beta = 7.29$, $SE = 1.47$, $p < .001$). MAILS shows an independent positive effect ($\beta = 2.27$, $SE = 0.72$, $p = .002$), indicating that higher AI literacy is also associated with more optimistic priors. NCS-6 has a negative but only marginally significant coefficient ($\beta = -1.99$, $SE = 1.05$, $p = .058$). Full results are reported in Appendix Table~\ref{tab:h5_appendix}. The appendix also visualizes prior distributions across tasks (Figure~\ref{fig:h5_violin}).

\begin{framed}
    \paragraph{Summary.}  
These findings support Hypothesis~5: dispositional trust in automation reliably predicts higher priors about AI accuracy. Moreover, AI literacy also shapes expectations, suggesting that both stable dispositions and learned knowledge influence how people approach new AI tasks—even before observing any outcomes.
\end{framed}


\begin{table}[t]
\centering
\caption{H5 results: Mixed-effects regressions of prior beliefs (0–100) on trust propensity (TiA). Standard errors in parentheses.}
\label{tab:h5_main}
\begin{tabular}{lcc}
\toprule
 & Model A & Model B \\
 & (TiA only) & (+ MAILS, NCS-6) \\
\midrule
TiA propensity (1–5) & 8.43*** & 7.29*** \\
 & (1.44) & (1.47) \\
MAILS (AI literacy) &  & 2.27** \\
 &  & (0.72) \\
NCS-6 (Need for cognition) &  & -1.99$^{\dagger}$ \\
 &  & (1.05) \\
\midrule
Task FE & Yes & Yes \\
Position FE & Yes & Yes \\
Participant RE & Yes & Yes \\
Observations & 720 & 720 \\
\bottomrule
\multicolumn{3}{l}{\footnotesize Notes: Dependent variable = prior belief about AI accuracy.}\\
\multicolumn{3}{l}{\footnotesize *** $p<0.001$, ** $p<0.01$, $^{\dagger} p<0.10$.}\\
\end{tabular}
\end{table}

\begin{table*}[t]
\centering
\caption{Summary of hypotheses (H1--H5) and outcomes.}
\label{tab:hypotheses_summary}
\begin{tabularx}{\linewidth}{lXc}
\toprule
\textbf{ID} & \textbf{Hypothesis} & \textbf{Outcome} \\
\midrule
H1 & Participants' priors of the AI system's accuracy are unrelated to experiences in prior, unrelated tasks. & Not supported \\
H2 & Participants rationally update their beliefs about the AI system's accuracy within each task following Bayes' rule. & Not supported (conservatism bias) \\
H3 & Participants with more positive beliefs about the AI system's accuracy delegate, \textit{ceteris paribus}, more. & Supported \\
H4 & Participants with less confidence in themselves delegate, \textit{ceteris paribus}, more. & Partially supported (only robust when beliefs are controlled) \\
H5 & Participants with a higher propensity to trust AI systems state higher AI accuracy priors before feedback. & Supported \\
\bottomrule
\end{tabularx}
\end{table*}

%% file: tex/6_Discussiom.tex
\section{Discussion and Implications}

Our study examined how people form, update, and act upon beliefs about AI systems across multiple tasks, combining preregistered hypotheses (H1–H5) with a large-scale experimental dataset (240 participants, 7,200 trials). Participants do not treat each new task as a clean slate, nor do they update beliefs in a fully rational Bayesian manner. Instead, they exhibit systematic spillovers, under-reaction, and reliance decisions shaped by both self-confidence and trust in AI. Below, we synthesize these findings and their implications.

\subsection{Belief stickiness across tasks (H1)}

Contrary to the rational benchmark that priors should reset independently in each new task, participants imported expectations from their most recent experience. A 10-point higher posterior in one task raised the prior in the next by roughly 3–4 points, even when domains were objectively unrelated. This pattern of \emph{belief stickiness} highlights a core risk for modern LLM-based systems, which increasingly bundle heterogeneous capabilities (e.g., grammar correction, question answering, retrieval, travel planning) under a single interface. If a user experiences the model as highly reliable in one domain—say, generating fluent text—they may carry that expectation into domains where reliability is lower, such as reasoning about factual knowledge or planning. Conversely, errors in one domain may undermine trust in otherwise competent features. 

Such spillovers mean that reliance on LLMs is not domain-isolated, but shaped by a global impression that can overgeneralize across contexts. Designers should therefore avoid assuming that users recalibrate their expectations for each task. Instead, systems may need to provide contextual performance cues (e.g., domain-specific reliability indicators or accuracy dashboards) and deliberately separate task framings (e.g., distinct affordances for creative writing vs.\ structured reasoning). Without such supports, belief inertia could either inflate misplaced confidence in fragile capabilities or unduly suppress use of well-calibrated ones.

These cross-task spillovers are indicative of users constructing a global mental model of a single AI agent, rather than forming domain-specific expectations. This aligns with \citet{kocielnik2019will} account of how expectations about imperfect AI systems shape satisfaction and acceptance, and with classic findings on the reliance gap in automation where users overgeneralize from salient errors \cite{dzindolet2003role, biswas2025mind}. By interpreting spillovers through this mental-model lens, we can see that users anchor on early impressions and apply them across tasks, producing durable belief inertia. This integration situates belief updating not just as a statistical phenomenon but as a process-level explanation for expectation formation and miscalibration in multipurpose AI systems.

This pattern is especially relevant as contemporary LLM platforms (e.g., ChatGPT, Gemini, Claude) unify many heterogeneous capabilities within a single agent frame, further reinforcing users’ tendency to form global expectations rather than domain-specific calibrations.

\subsection{Bounded updating within tasks (H2)}
Within tasks, participants updated in the correct direction but only half as much as Bayes’ rule prescribes. This \emph{conservatism bias} echoes decades of cognitive psychology work on belief updating \cite{Edwards1968, Tversky1974}, but here it emerges in an applied Human–AI interaction setting. The implication is that users may under-react to both AI successes and failures, converging too slowly to the AI’s true accuracy. 

For LLM-based applications, this dynamic is consequential. A user who sees a model hallucinate once may continue to treat it as less reliable than it actually is, even after many correct responses. Conversely, a string of correct answers may not fully restore confidence if users are reluctant to update. This helps explain why users often oscillate between over- and under-reliance on LLMs despite extended exposure. 

Designers can counteract such inertia by making evidence integration more salient: surfacing transparent performance histories (e.g., “8/10 factual responses verified this session”), clarifying how much evidence a user has actually observed (e.g., “so far based on 3 examples”), or providing calibrated self-assessments from the model itself. Without such supports, users may remain miscalibrated even after long-term interaction, leading to brittle trust in everyday deployments like search, coding assistants, or productivity tools.

Interpreting these bounded updates through established behavioral mechanisms clarifies the cognitive processes underlying our results. Conservative adjustments align with selective‐attention models, such as sparsity‐based bounded rationality \cite{gabaix2014sparsity}, where users encode only sparse or partial evidence, thereby dampening responsiveness. The persistence of initial beliefs is consistent with anchoring and sequential belief–averaging dynamics \cite{hogarth1992order}. Taken together, these patterns suggest that users form and revise their mental models of AI under cognitive resource constraints, producing belief paths that systematically diverge from Bayesian predictions. These mechanism level insights clarify what our statistical effects reflect in terms of cognitive process and provide a bridge to broader HCI constructs—such as miscalibration, over-reliance, and system abandonment—where similar biases have been observed.

\subsection{Beliefs as the dominant driver of delegation (H3)}

Across models, beliefs in AI accuracy emerged as the strongest predictor of delegation. Participants who believed the system would be correct were substantially more likely to hand decisions to it, with odds of delegation rising steeply as beliefs increased. Importantly, this effect was strongest in grammar, attenuated in travel, and weakest in VQA—consistent with ceiling effects in perceptual tasks where delegation was already high. 

For LLM-based systems, this finding underscores that delegation is driven less by objective accuracy and more by users’ \emph{subjective beliefs} about accuracy. 
This pattern reflects classic reliance calibration dynamics, where subjective expectations rather than ground truth performance govern acceptance of automation \cite{dzindolet2003role, lee2004trust, hoff2015trust}. A user who perceives the model as highly competent at drafting an email may also be more willing to let it generate a trip itinerary or summarize a paper, even if its actual performance varies dramatically across these domains. Conversely, if early interactions lower beliefs, users may avoid delegation even in domains where the model is strong. 

This has two practical implications. First, calibrating beliefs may be more important for shaping reliance than marginal accuracy gains. Transparent uncertainty displays, error exemplars, or domain-specific performance feedback can help align user beliefs with true capability. Second, because beliefs generalize unevenly across tasks, bundling multiple LLM functions under one interface risks producing both over-reliance (delegating where accuracy is weak) and under-reliance (ignoring strong capabilities). Designing interventions that localize beliefs to the task at hand—through contextual affordances or scoped feedback—may be critical to fostering calibrated reliance in multi-tool LLM environments.

Since delegation decisions are downstream of users’ beliefs, the same mechanisms that shape belief updating—such as anchoring, selective attention, and confirmation-driven expectation maintenance—also propagate into reliance choices. These processes help explain why users may continue to delegate despite inconsistent evidence, or refrain from delegation even when the AI is demonstrably accurate.

\subsection{Confidence and the Self vs. AI trade-off (H4)}

Self-confidence also shaped delegation, though more weakly and contingently than beliefs. When considered alone, the effect of confidence was mixed; but once beliefs were accounted for, confidence robustly reduced delegation. Participants who felt more capable preferred to rely on themselves, while those with lower confidence were more willing to hand decisions to the AI. Beliefs nonetheless exerted the stronger influence. 

This dynamic is especially relevant for LLM deployments. A confident writer may use an LLM as a drafting assistant but ultimately choose to edit and finalize text themselves, while a less confident writer may defer wholesale to the model. Similarly, novice coders often rely heavily on code completions, whereas experts treat them as suggestions to be verified. In both cases, delegation reflects a trade-off between trust in one’s own ability and belief in the model’s competence. 

For design, this means that reliance is not determined by system performance alone but by how capable users feel in a given domain. If interfaces erode self-confidence—through opaque feedback, over-assertive suggestions, or framing the model as an “expert”—they may inadvertently push users toward over-reliance. Conversely, designs that scaffold user competence (e.g., explain why a completion works, or encourage reflection before accepting a generated plan) can support healthier calibration. Addressing this self–AI balance is essential to prevent both disuse among confident experts and over-reliance among novices.

\subsection{Dispositional trust and initial expectations (H5)}

Finally, dispositional trust in automation (TiA) and, to a lesser extent, AI literacy shaped participants’ initial priors before any feedback. Individuals who generally trusted automation entered each task with systematically higher expectations of the AI’s accuracy, while those higher in AI literacy also began more optimistic. These results show that reliance trajectories are not formed from scratch but are colored by stable traits and prior knowledge. 

For LLMs, this means that two users encountering the same model—say, a search assistant or a coding companion—may start with very different baseline expectations. A highly trusting user may initially delegate generously, interpreting early mistakes as anomalies, while a skeptical user may under-delegate even when the model is competent. This divergence raises design challenges: onboarding mechanisms must actively recalibrate priors, rather than assuming users converge naturally. Providing representative performance demonstrations, calibrated self-assessments, or task-specific “warm-up” examples could help align initial expectations across diverse user populations.

Following \citet{tyack2024self} call for stronger theoretical grounding in HCI, we position belief updating as a mechanism that explains how users come to hold the expectations that drive reliance, calibration, and satisfaction. Whereas prior HCI work has emphasized constructs such as trust, inaccuracy tolerance, and mental-model correctness, our approach makes explicit the dynamic processes that produce these states. By showing when users under-react, overgeneralize, or anchor on early impressions, we expose the cognitive mechanisms behind reliance gaps, discrimination gaps, and abandonment patterns reported in prior research. This integration clarifies our contribution: belief updating is not only a descriptive tool but a theoretical framework for understanding the formation and evolution of user expectations across multipurpose AI systems, offering new leverage for designing systems that better support appropriate and well-calibrated reliance decisions. This highlights the importance of onboarding mechanisms that explicitly calibrate initial expectations, rather than assuming that users will naturally converge to accurate priors through use alone.

\subsection{Caveats, Limitations, and Future Directions}

While many contemporary AI systems support iterative, conversational forms of interaction, our study intentionally adopts a one-shot delegation structure. This choice allows each trial to correspond to a single reliance decision grounded in clearly interpretable evidence. In more open-ended settings—such as conversational co-creation or writing assistance—users often engage in successive refinement, verification, and partial editing of AI-generated content. Our framework is conceptually compatible with such settings, as additional AI turns can be viewed as additional pieces of evidence; however, the unit of evidence and the unit of delegation become substantially noisier to define in multi-turn interaction. Iterative interfaces introduce confounds such as exploration strategies, prompt-refinement skills, and mixed-initiative control dynamics, which obscure the mechanistic relationship between belief updating and reliance. For this reason, we treat our controlled, one-shot paradigm as a mechanism-oriented first step that isolates how beliefs and confidence shape reliance. Extending this framework to richer, multi-turn human–AI collaboration remains an important avenue for future HCI research.

Our study provides immediate corrective feedback after every trial, showing participants the ground truth, their own answer, and the AI’s answer. While not all real-world AI interactions provide verifiable outcomes, many widely used AI applications do offer rapid or instantaneous correctness signals. Examples include coding assistants (where compilation or runtime outputs reveal correctness), grammar and writing tools (where users can verify corrected sentences), search and question-answering with citations, fact-checking workflows, content moderation labeling interfaces, and tasks with objective outputs such as math, data cleaning, or form completion. In these settings, users routinely observe whether the AI was correct, even when the system itself does not supply explicit feedback.

Immediate feedback in our design is therefore not meant to mimic all AI use cases, but to isolate belief-updating dynamics by providing clean, unambiguous evidence. Without ground-truth feedback, belief updating becomes difficult to identify because changes in beliefs are confounded with uncertainty about task difficulty or users’ own accuracy. Our approach follows long-standing calibration and trust-in-automation paradigms, which require observable outcomes to benchmark deviations from normative updating. Future work should examine belief formation in tasks where correctness is subjective or unverifiable, and where users must actively decide whether, when, and how to verify AI output.

Our study provides systematic evidence for how beliefs, confidence, and dispositional traits shape reliance on AI, but several limitations qualify the findings. First, the experimental tasks—grammar checking, travel planning, and visual question answering—capture important but bounded slices of LLM functionality. While they reflect core categories of LLM use (text editing, planning, and perception–reasoning), future work should examine whether the same patterns hold in other capabilities such as coding, scientific summarization, or multimodal creative work. Extending the paradigm to more diverse domains will test the generality of belief spillovers and bounded updating.

Second, task accuracy was experimentally fixed at predetermined levels (low for grammar, moderate for travel, high for VQA). This choice allowed clean comparisons across conditions but departs from the variable performance that users encounter in real-world LLMs, where accuracy fluctuates not only by domain but also by prompt formulation, context, or user expertise. Future work should explore dynamic or stochastic accuracy trajectories to capture how people respond to uncertainty and inconsistency in AI performance.

We also note that the AI’s answers were pre-scripted rather than generated by a live model. This choice is methodologically necessary for isolating belief updating: real LLMs exhibit stochastic, prompt-sensitive accuracy that cannot be held constant across trials. Pre-scripted responses allow each trial to correspond to a well-defined likelihood signal (correct vs. incorrect). This approach is consistent with classic calibration and trust-in-automation paradigms and mirrors many real-world settings with objective correctness signals (e.g., code execution, form validation, grammar correction). Future work should examine belief formation in tasks with subjective or unverifiable correctness, where users must actively decide whether and how to verify AI output.

Third, task difficulty was not equated across domains. Grammar, travel, and VQA differ in intrinsic complexity and in how confident participants may feel about their own baseline competence. This asymmetry likely influenced both self-confidence and delegation tendencies. For example, grammar may have felt more tractable for many users, amplifying the role of beliefs in delegation, whereas travel and VQA involved higher uncertainty and led to ceiling effects. Future studies should systematically manipulate task difficulty to disentangle whether delegation patterns stem from system performance, user confidence, or task complexity.

%% file: tex/7_Conclusion.tex
\section{Conclusions}
In sum, our findings draw nuances to the rationalist picture of human–AI reliance. Users do not reset beliefs across tasks, do not update fully within tasks, and delegate based on subjective beliefs and confidence rather than objective accuracy alone. Yet these deviations from rationality are systematic and interpretable. By surfacing the layered structure of reliance—cross-task spillovers, bounded updates, belief-driven delegation, and dispositional baselines—we provide actionable insights for building AI systems that foster calibrated trust and safe delegation. For HCI, this highlights the need to design not just for accuracy, but for the human cognitive and dispositional processes that govern how AI is understood, trusted, and used.

%% file: tex/8_Appendix.tex
\newpage
\section{Appendix}

\begin{table}[H]
\centering
\caption{Appendix H1. Pooled OLS with participant-clustered SEs.}
\label{app:h1_ols}
\begin{tabular}{lcc}
\toprule
 & Coef. & SE \\
\midrule
Previous posterior ($bT_{\text{prev}}$) & 0.432*** & (0.045) \\
Task FE & Yes &  \\
Position FE & Yes &  \\
Clustered SEs (participant) & Yes & \\
\midrule
Observations & 480 & \\
$R^2$ & 0.312 & \\
\bottomrule
\multicolumn{2}{l}{\footnotesize Dependent variable is $b0_{\text{curr}}$.}\\
\multicolumn{2}{l}{\footnotesize *** $p<0.001$.}\\
\end{tabular}
\end{table}

\begin{table}[H]
\centering
\caption{Appendix H1. OLS with previous objective AI performance.}
\label{app:h1_ols_robust}
\begin{tabular}{lcc}
\toprule
 & Coef. & SE \\
\midrule
Previous posterior ($bT_{\text{prev}}$) & 0.425*** & (0.046) \\
Previous objective AI perf. & 0.018 & (0.024) \\
Task FE & Yes &  \\
Position FE & Yes &  \\
Clustered SEs (participant) & Yes & \\
\midrule
Observations & 480 & \\
$R^2$ & 0.313 & \\
\bottomrule
\multicolumn{2}{l}{\footnotesize Dependent variable is $b0_{\text{curr}}$.}\\
\multicolumn{2}{l}{\footnotesize *** $p<0.001$. $p=.459$ for objective perf.}\\
\end{tabular}
\end{table}

\begin{table*}[t]
\centering
\caption{Appendix H1. Mixed-effects regression with objective AI performance only.}
\label{app:h1_mixed_prevObj}
\begin{tabular}{lcc}
\toprule
 & Coef. & SE \\
\midrule
Previous objective AI perf. & 0.088*** & (0.023) \\
Task FE & Yes &  \\
Position FE & Yes &  \\
Random intercept (participant) & Yes & \\
\midrule
Observations & 480 & \\
Log-likelihood & -2048.8 & \\
\bottomrule
\multicolumn{2}{l}{\footnotesize Dependent variable is $b0_{\text{curr}}$.}\\
\multicolumn{2}{l}{\footnotesize *** $p<0.001$.}\\
\end{tabular}
\end{table*}

\begin{table*}[b]
\centering
\caption{Appendix H1. Participant fixed-effects (demeaned OLS).}
\label{app:h1_participantFE}
\begin{tabular}{lcc}
\toprule
 & Coef. & SE \\
\midrule
Previous posterior (demeaned) & 0.041 & (0.050) \\
Task FE & Yes &  \\
Position FE & Yes &  \\
Participant FE & Yes & \\
\midrule
Observations & 480 & \\
$R^2$ & 0.063 & \\
\bottomrule
\multicolumn{2}{l}{\footnotesize DV is $b0_{\text{curr}}$ (demeaned within participant).}\\
\multicolumn{2}{l}{\footnotesize n.s. effect: $p=.414$.}\\
\end{tabular}
\end{table*}

\begin{table*}[t]
\centering
\caption{Appendix H1. Belief-change model ($b0_{\text{curr}} - bT_{\text{prev}}$).}
\label{app:h1_change}
\begin{tabular}{lcc}
\toprule
 & Coef. & SE \\
\midrule
Intercept (overall reset) & -2.56 & (2.02) \\
Travel vs.\ grammar & 7.36** & (2.48) \\
VQA vs.\ grammar & 9.03*** & (2.47) \\
Task position (3 vs.\ 2) & 3.36 & (2.01) \\
\midrule
Observations & 480 & \\
Log-likelihood & -2157.8 & \\
\bottomrule
\multicolumn{2}{l}{\footnotesize DV is difference between prior in current task and posterior in previous task.}\\
\multicolumn{2}{l}{\footnotesize ** $p<0.01$, *** $p<0.001$.}\\
\end{tabular}
\end{table*}

\begin{table*}[b]
\centering
\caption{Appendix H1. Mixed-effects regression of current-task priors ($b0_{\text{curr}}$) with pre-survey controls (MAILS, NCS-6).}
\label{app:h1_controls}
\begin{tabular}{lcc}
\toprule
 & Coef. & SE \\
\midrule
Intercept & 40.03*** & (5.73) \\
Task: Travel & -0.48 & (1.68) \\
Task: VQA & -2.54 & (1.73) \\
Position: 3 & 2.42* & (1.23) \\
Previous posterior ($bT_{\text{prev}}$) & 0.297*** & (0.040) \\
MAILS (AI literacy) & 2.23** & (0.67) \\
NCS-6 (Need for cognition) & -0.83 & (0.99) \\
\midrule
Task FE & Yes \\
Position FE & Yes \\
Participant RE & Yes \\
Observations & 480 \\
Log-likelihood & -2017.2 \\
\bottomrule
\multicolumn{3}{l}{\footnotesize Dependent variable = prior belief (0--100).}\\
\multicolumn{3}{l}{\footnotesize *** $p<0.001$, ** $p<0.01$, * $p<0.05$.}\\
\end{tabular}
\end{table*}



\begin{table*}[t]
\centering
\caption{Appendix H2. Regression of observed belief updates on Bayesian normative steps, by counterfactual policy. Clustered SEs at participant level.}
\label{app:h2_robust}
\begin{tabular}{lcccc}
\toprule
Policy & $N$ updates & $\hat{\sigma}$ & SE & $p$-value \\
\midrule
Strict (CF only)       & 3,430 & 0.52 & 0.06 & $<.001$ \\
Lenient (CF only)      & 5,670 & 0.47 & 0.05 & $<.001$ \\
Hybrid (fallback $S=5$)  & 6,560 & 0.47 & 0.04 & $<.001$ \\
Hybrid (fallback $S=10$) & 6,560 & 0.49 & 0.05 & $<.001$ \\
Hybrid (fallback $S=20$) & 6,560 & 0.49 & 0.05 & $<.001$ \\
\bottomrule
\end{tabular}
\end{table*}


\begin{figure*}[b]
  \centering

  \begin{subfigure}{0.48\linewidth}
    \includegraphics[width=\linewidth]{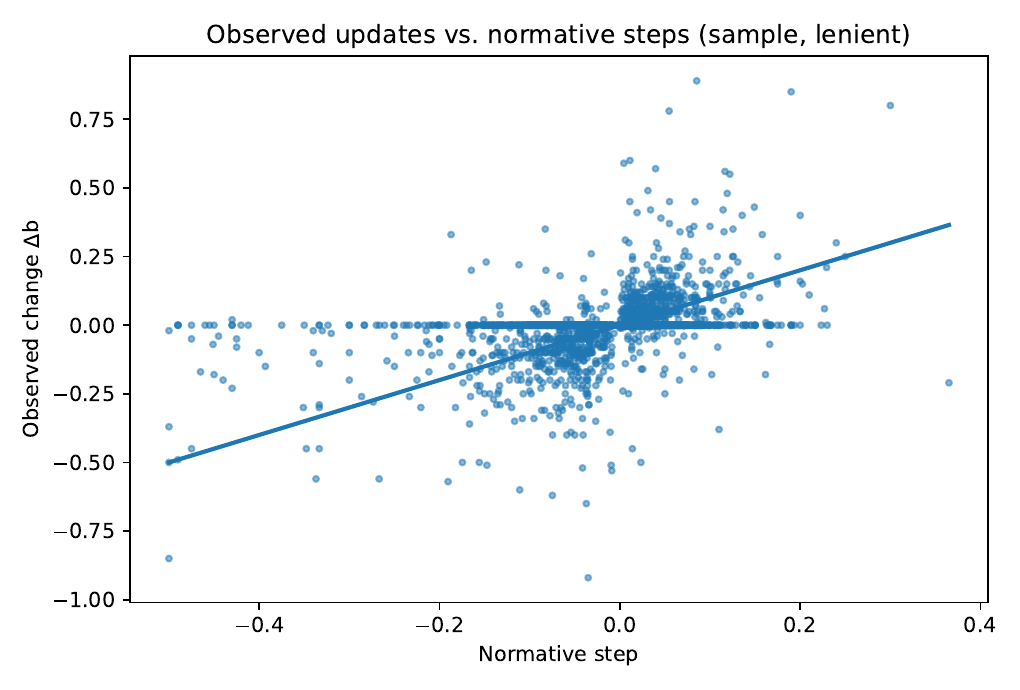}
    \caption{Lenient}
  \end{subfigure}
  \hfill
  \begin{subfigure}{0.48\linewidth}
    \includegraphics[width=\linewidth]{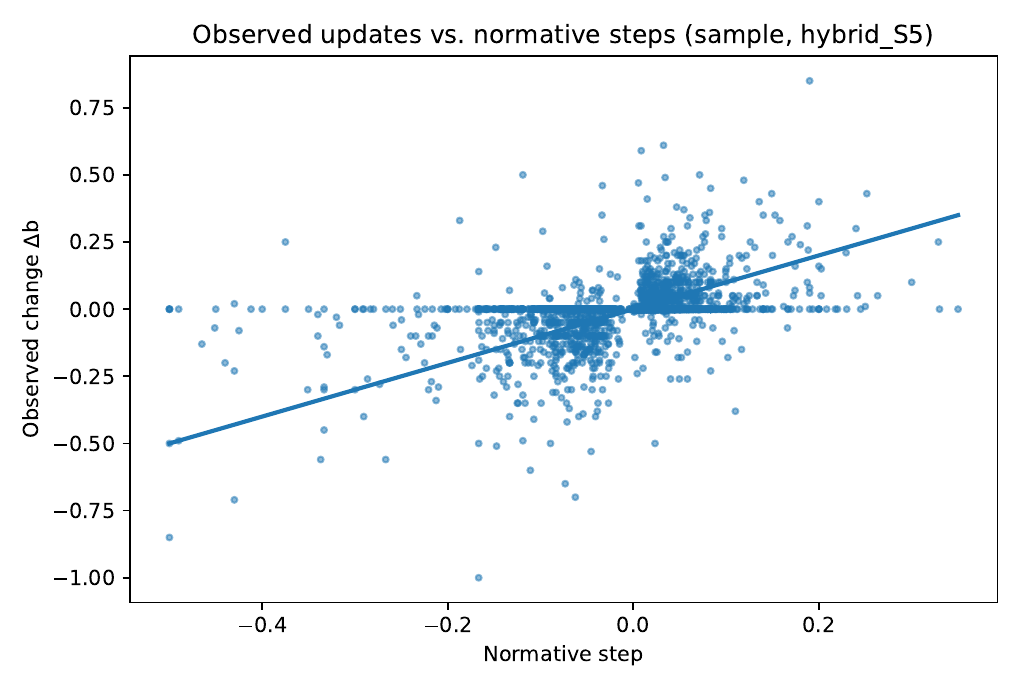}
    \caption{Hybrid $S=5$}
  \end{subfigure}

  \par\medskip

  \begin{subfigure}{0.48\linewidth}
    \includegraphics[width=\linewidth]{fig/H2/h2_scatter_hybrid_S10.pdf}
    \caption{Hybrid $S=10$}
  \end{subfigure}
  \hfill
  \begin{subfigure}{0.48\linewidth}
    \includegraphics[width=\linewidth]{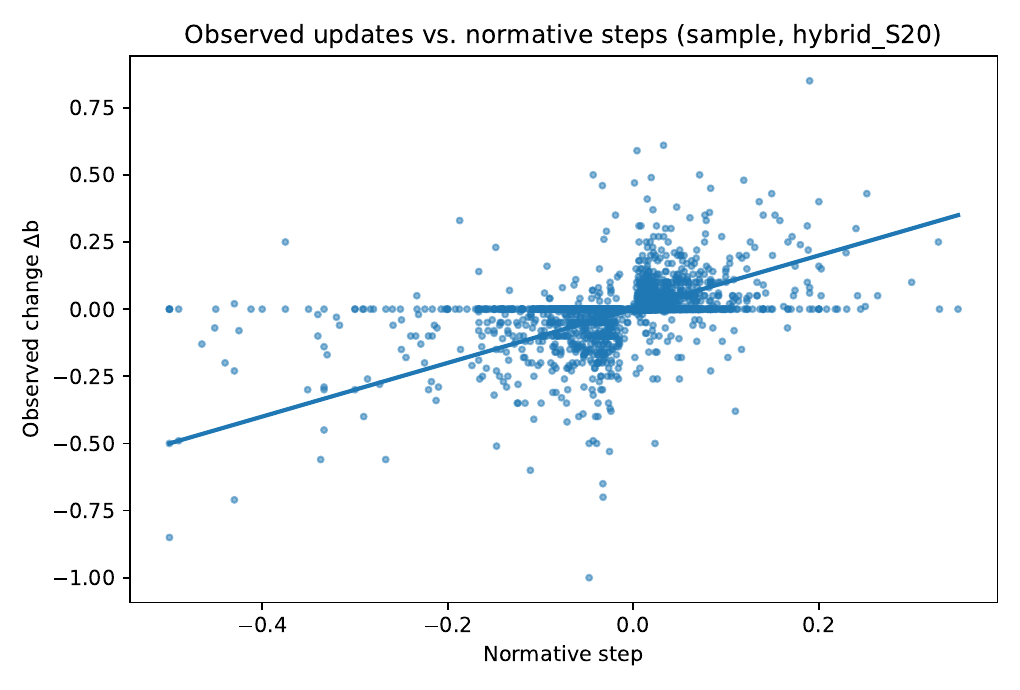}
    \caption{Hybrid $S=20$}
  \end{subfigure}

  \caption{Appendix H2. Observed vs.\ normative belief updates under alternative CF policies. 
  All slopes $\hat{\sigma}\approx 0.47$–0.49, indicating conservatism.}
  \label{fig:h2_scatter_appendix}
\end{figure*}


\begin{figure*}[t]
  \centering
  \begin{subfigure}{.48\linewidth}
    \includegraphics[width=\linewidth]{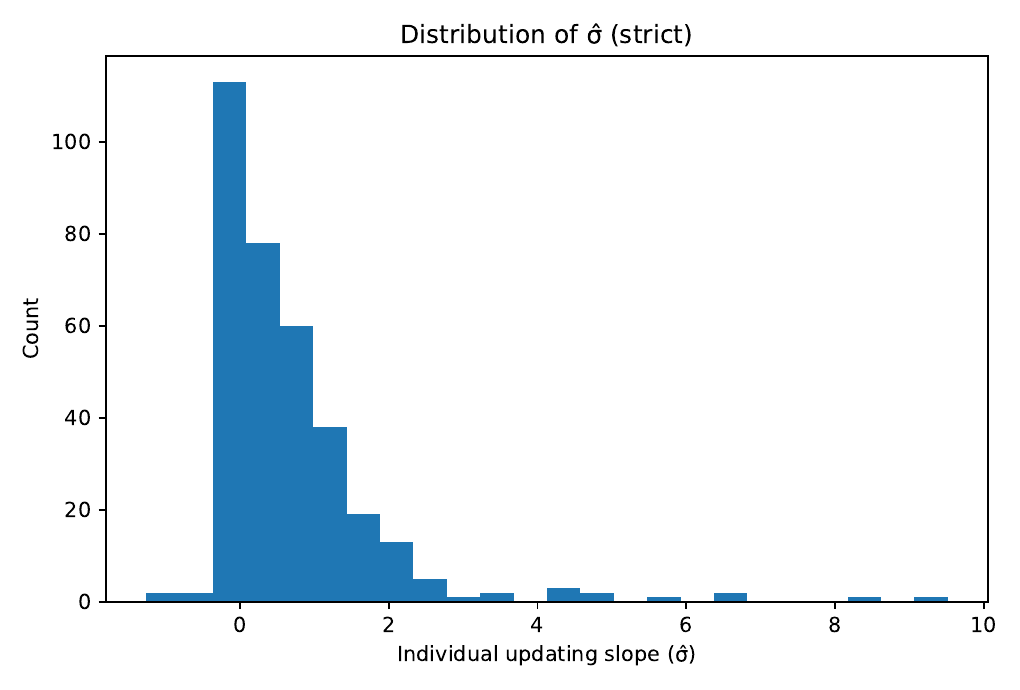}
    \caption{Strict}
  \end{subfigure}
  \hfill
  \begin{subfigure}{.48\linewidth}
    \includegraphics[width=\linewidth]{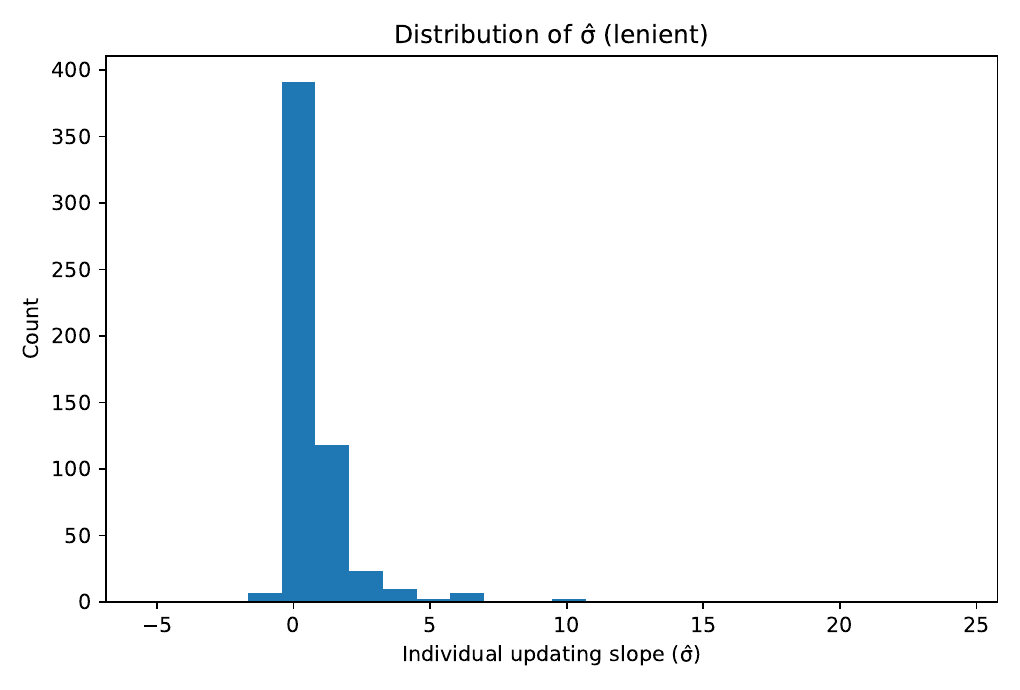}
    \caption{Lenient}
  \end{subfigure}
  
  
  \begin{subfigure}{.48\linewidth}
    \includegraphics[width=\linewidth]{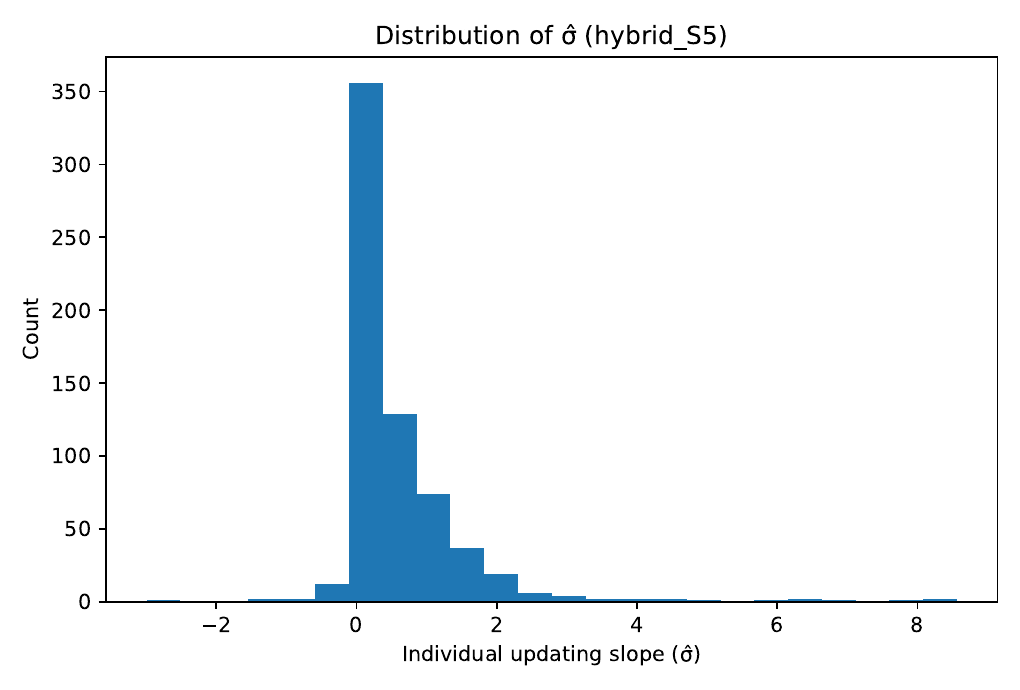}
    \caption{Hybrid $S=5$}
  \end{subfigure}
  \hfill
  \begin{subfigure}{.48\linewidth}
    \includegraphics[width=\linewidth]{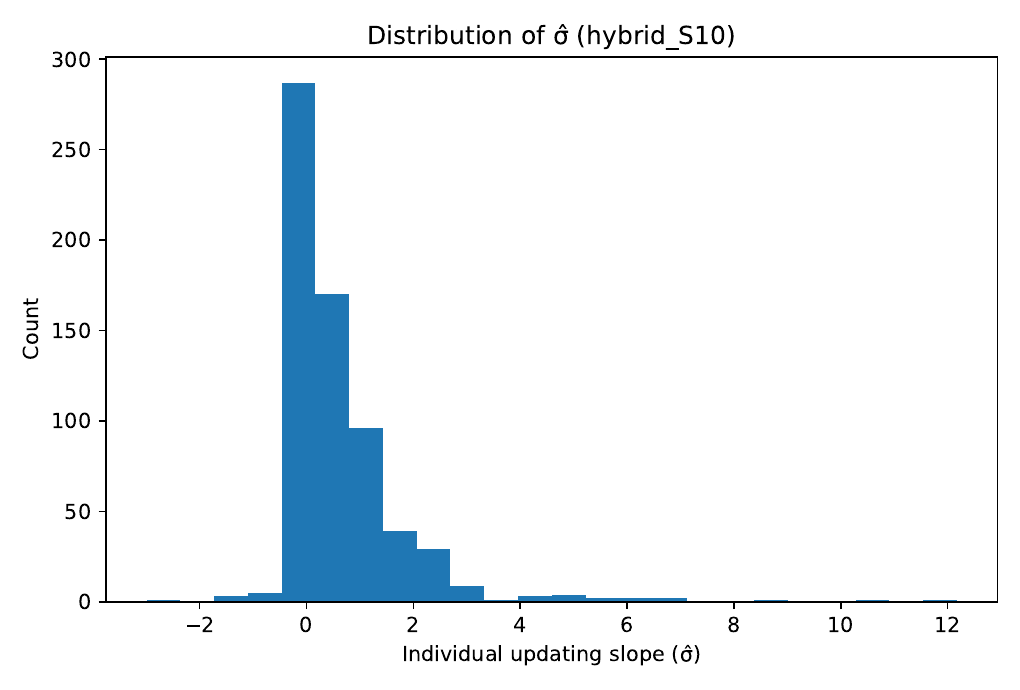}
    \caption{Hybrid $S=10$}
  \end{subfigure}
  \hfill
  \begin{subfigure}{.48\linewidth}
    \includegraphics[width=\linewidth]{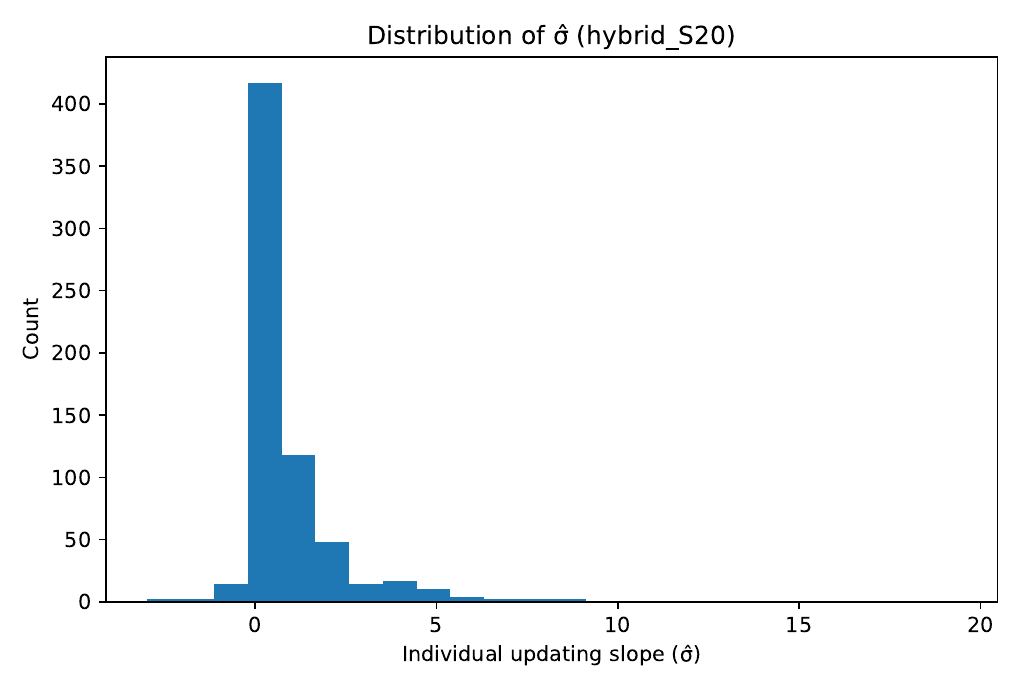}
    \caption{Hybrid $S=20$}
  \end{subfigure}
  \caption{Appendix H2. Distribution of individual updating slopes ($\hat{\sigma}$) across CF policies. Most participants under-react relative to the Bayesian benchmark ($\sigma=1$).}
  \label{fig:h2_hist_appendix}
\end{figure*}

\begin{table*}[H]
\centering
\caption{Appendix H2. Task-wise regressions of observed belief updates on Bayesian normative steps under the hybrid counterfactual policy ($S=10$). Clustered SEs at the participant level.}
\label{app:h2_taskwise}
\begin{tabular}{lcccc}
\toprule
Task & $N$ updates & $\hat{\sigma}$ & SE & $p$-value \\
\midrule
Grammar & 2,230 & 0.46 & 0.06 & $<.001$ \\
Travel  & 2,150 & 0.50 & 0.05 & $<.001$ \\
VQA     & 2,180 & 0.61 & 0.09 & $<.001$ \\
\bottomrule
\end{tabular}
\end{table*}


\begin{table*}[t]
\centering
\caption{Appendix H3. Logistic regression of delegation on lagged belief (base models). Dependent variable: delegate AI (1=yes).}
\label{app:h3_base}
\begin{tabular}{lcc}
\toprule
 & GLM (clustered SE) & GEE (exchangeable) \\
\midrule
Lagged belief ($b_{t-1}$) & 1.819*** (0.321) & 1.670*** (0.315) \\
Travel task (ref = Grammar) & 0.811*** (0.177) & 0.782*** (0.180) \\
VQA task (ref = Grammar) & 1.177*** (0.177) & 1.154*** (0.183) \\
Trial FE & Yes & Yes \\
Task position FE & Yes & Yes \\
\midrule
Observations & 6,720 & 6,720 \\
\bottomrule
\multicolumn{3}{l}{\footnotesize Notes: Logistic regression coefficients; SE in parentheses. *** $p<0.001$.} \\
\end{tabular}
\end{table*}

\begin{table*}[b]
\centering
\caption{Appendix H3. Logistic regression with pre-survey controls (belief-only specification).}
\label{app:h3_controls}
\begin{tabular}{lc}
\toprule
 & GLM (clustered SE) \\
\midrule
Lagged belief ($b_{t-1}$) & 1.756*** (0.328) \\
Travel task (ref = Grammar) & 0.793*** (0.181) \\
VQA task (ref = Grammar) & 1.163*** (0.182) \\
TiA (propensity) & 0.292* (0.145) \\
MAILS (AI literacy) & -0.084 (0.116) \\
NCS-6 (cognition) & 0.039 (0.104) \\
Trial FE, Task position FE & Yes \\
\midrule
Observations & 6,240 \\
\bottomrule
\multicolumn{2}{l}{\footnotesize Notes: Logistic regression coefficients; SE in parentheses. * $p<0.05$, *** $p<0.001$.} \\
\end{tabular}
\end{table*}

\begin{table*}[t]
\centering
\caption{Appendix H3. Logistic regression with belief × task interaction.}
\label{app:h3_interaction}
\begin{tabular}{lc}
\toprule
 & GLM (clustered SE) \\
\midrule
Lagged belief ($b_{t-1}$) & 2.104*** (0.406) \\
$\;\;\times$ Travel & -0.681* (0.312) \\
$\;\;\times$ VQA & -0.927** (0.295) \\
Travel task (ref = Grammar) & 0.858*** (0.186) \\
VQA task (ref = Grammar) & 1.194*** (0.188) \\
Trial FE, Task position FE & Yes \\
\midrule
Observations & 6,720 \\
\bottomrule
\multicolumn{2}{l}{\footnotesize Notes: Logistic regression coefficients; SE in parentheses. * $p<0.05$, ** $p<0.01$, *** $p<0.001$.} \\
\end{tabular}
\end{table*}


\begin{table*}[b]
\centering
\caption{Appendix H4. Conf-only models predicting delegation. Dependent variable: $delegate_{it}$ (AI chosen).}
\label{app:h4_confonly}
\begin{tabular}{lcc}
\toprule
 & GLM (clustered SE) & GEE (exchangeable) \\
\midrule
Lagged confidence ($conf_{t-1}$) & -1.12*** & -1.08** \\
 & (0.35) & (0.32) \\
Task FE & Yes & Yes \\
Trial FE & Yes & Yes \\
Participant RE & -- & Yes \\
\midrule
Observations & 7,200 & 7,200 \\
\bottomrule
\multicolumn{3}{l}{\footnotesize Notes: Coefficients are log-odds. Standard errors in parentheses.}\\
\multicolumn{3}{l}{\footnotesize ** $p<0.01$, *** $p<0.001$.}\\
\end{tabular}
\end{table*}

\begin{table*}[t]
\centering
\caption{Appendix H4. Joint belief + confidence model. Dependent variable: $delegate_{it}$.}
\label{app:h4_joint}
\begin{tabular}{lcc}
\toprule
 & GLM (clustered SE) & GEE (exchangeable) \\
\midrule
Lagged confidence ($conf_{t-1}$) & -1.54*** & -1.47*** \\
 & (0.34) & (0.33) \\
Lagged belief ($b_{t-1}$) & 1.92*** & 1.85*** \\
 & (0.32) & (0.31) \\
Task FE & Yes & Yes \\
Trial FE & Yes & Yes \\
Participant RE & -- & Yes \\
\midrule
Observations & 7,200 & 7,200 \\
\bottomrule
\multicolumn{3}{l}{\footnotesize Notes: Coefficients are log-odds. Standard errors in parentheses.}\\
\multicolumn{3}{l}{\footnotesize *** $p<0.001$.}\\
\end{tabular}
\end{table*}

\begin{table*}[b]
\centering
\caption{Appendix H4. Pre-only robustness (trial 1). Dependent variable: $delegate_{i1}$.}
\label{app:h4_preonly}
\begin{tabular}{lc}
\toprule
 & GLM (clustered SE) \\
\midrule
Pre-task confidence & -0.23 \\
 & (0.62) \\
Task FE & Yes \\
Participant RE & -- \\
\midrule
Observations & 720 \\
\bottomrule
\multicolumn{2}{l}{\footnotesize Notes: Coefficients are log-odds. Standard errors in parentheses.}\\
\end{tabular}
\end{table*}

\begin{table*}[t]
\centering
\caption{Appendix H5. Robustness summary: priors regressed on TiA with controls (recap of Model B).}
\label{tab:h5_appendix}
\begin{tabular}{lc}
\toprule
 & Model B (with MAILS, NCS-6) \\
\midrule
TiA propensity (1--5) & 7.29*** \\
 & (1.47) \\
MAILS (AI literacy) & 2.27** \\
 & (0.72) \\
NCS-6 (Need for cognition) & -1.99$^{\dagger}$ \\
 & (1.05) \\
\midrule
Task FE, Position FE, Participant RE & Yes \\
Observations & 720 \\
\bottomrule
\multicolumn{1}{l}{\footnotesize Dependent variable = prior belief (0--100). Standard errors in parentheses.}\\
\multicolumn{1}{l}{\footnotesize *** $p<0.001$, ** $p<0.01$, $^{\dagger} p<0.10$.}
\end{tabular}
\end{table*}

\begin{figure*}[b]
  \centering
  \includegraphics[width=0.7\linewidth]{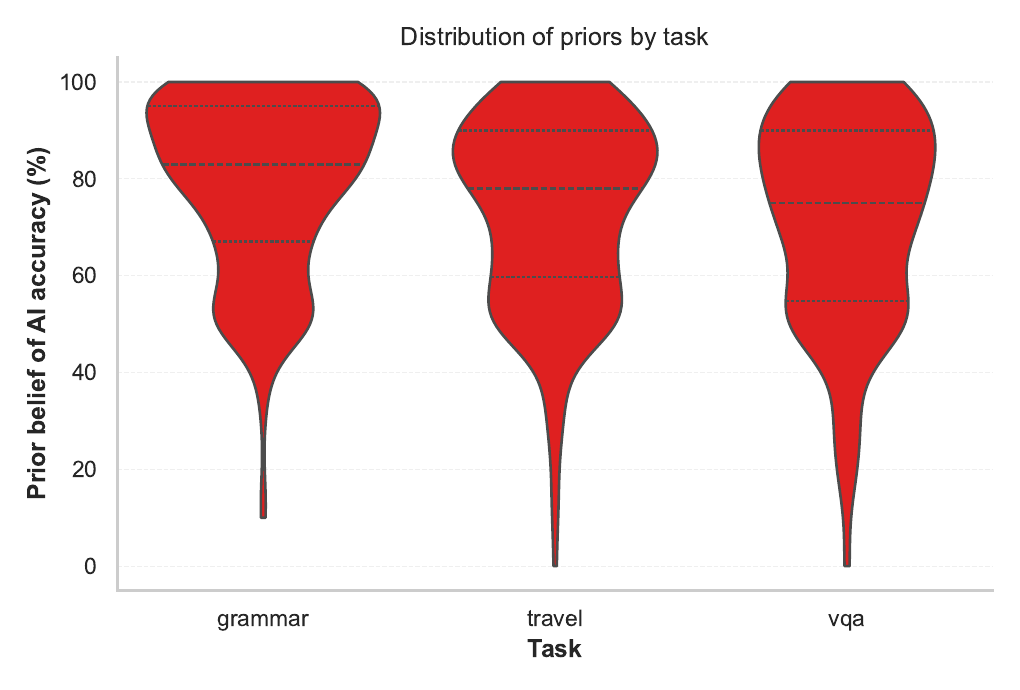}
  \caption{Appendix H5. Distributions of trial-1 priors (0--100) by task (grammar, travel, VQA). Points show medians; boxes show IQR.}
  \label{fig:h5_violin}
\end{figure*}